\documentclass[aps, pre, twocolumn, floatfix]{revtex4-1}
\usepackage{graphicx}
\usepackage{amssymb, amsfonts, amsmath}
\usepackage{epsf}
\usepackage{subfigure}
\usepackage{epstopdf}
\usepackage{txfonts}
\usepackage{mathrsfs}
\usepackage{chemarrow}

\usepackage{color}

\newcommand{\be}{\begin{equation}}
\newcommand{\ee}{\end{equation}}
\newcommand{\bea}{\begin{eqnarray}}
\newcommand{\eea}{\end{eqnarray}}

\begin{document}

\title{Stochastic approach and fluctuation theorem for charge transport in diodes}

\author{Jiayin Gu}
\email{\tt jiaygu@ulb.ac.be}
\author{Pierre Gaspard}
\email{\tt gaspard@ulb.ac.be}
\affiliation{Center for Nonlinear Phenomena and Complex Systems, Universit\'e Libre de Bruxelles (U.L.B.), Campus Plaine, Code Postal 231, B-1050 Brussels, Belgium}

\begin{abstract}
A stochastic approach for charge transport in diodes is developed in consistency with the laws of electricity, thermodynamics, and microreversibility.  In this approach, the electron and hole densities are ruled by diffusion-reaction stochastic partial differential equations and the electric field generated by the charges is determined with the Poisson equation.  These equations are discretized in space for the numerical simulations of the mean density profiles, the mean electric potential, and the current-voltage characteristics.  Moreover, the full counting statistics of the carrier current and the measured total current including the contribution of the displacement current are investigated.  On the basis of local detailed balance, the fluctuation theorem is shown to hold for both currents.
\end{abstract}

\maketitle

\section{Introduction}

Diodes are nonlinear electronic devices used for current rectification.  The core of a diode is a junction between  semiconductors doped with positively and negatively charged impurities \cite{AM76}.  As shown by Shockley \cite{S49}, the electric potential barrier generated at the junction induces the highly nonlinear and asymmetric current-voltage characteristic at the basis of rectification.

Besides their great technological interest, diodes can be used to address the fundamental issue of microreversibility in nonequilibrium statistical mechanics.  In the regime of linear response, the Onsager-Casimir reciprocal relations are well established consequences of microreversibility \cite{O31a,O31b,C45}.  However, their domain of validity is much restricted in highly nonlinear electronic devices.  Recent advances have shown that time-reversal symmetry relations also hold in the regimes of nonlinear response arbitrarily away from equilibrium.  These relations, called fluctuation theorems \cite{G96,ES02,EHM09,CHT11,J11,S12,G13}, concern the fluctuations of nonequilibrium work, heat, or currents in systems driven out of equilibrium by time-dependent forces or by reservoirs of particles or energy at their boundaries. The fluctuations satisfying these theorems are generated by the erratic movements of the particles composing matter and they manifest themselves as noises such as the Johnson-Nyquist and shot noises \cite{J28,N28,BB00}.  They are thus ruled by statistical laws obeying microreversibility.  In particular, the fluctuation theorem for currents implies the Onsager-Casimir reciprocal relations, as well as higher-order generalizations to the regime of nonlinear response \cite{G13,AG04,AG07JSM,HPPG11}.

In electric systems, fluctuation theorems for nonequilibrium work and heat have been theoretically and experimentally investigated in linear $RC$ circuits \cite{vZCC04,GC05,JGC08}, leaving open the study of current fluctuations in nonlinear circuits.  This is motivating the need to develop a stochastic approach and to establish the fluctuation theorem for charge transport in diodes,  consistently with the laws of electricity, thermodynamics, and microreversibility \cite{AG09}. Stochastic models have already been proposed for the random number of charges crossing a diode \cite{H68,LH10} or mesoscopic junctions \cite{AWBMJ91,AG06}.  

Here, our purpose is to develop a spatially extended stochastic description of the diode junction.  Since electron-hole pairs are generated and recombined in semiconducting $p$-$n$ junctions, the approach is based on diffusion-reaction stochastic partial differential equations for the charge carrier densities.  These balance equations are coupled to the Poisson equation for the electric potential in the quasi-static limit of Maxwell's equations \cite{J99}.  Therefore, the electric current fluctuations are deeply influenced by the long-ranged Coulomb interaction and the displacement current should be included to determine the measured  current \cite{BB00,AG09}.  In addition, the stochastic process satisfies local detailed balance as a consequence of microreversibility.  In this framework, the fluctuation theorem is shown to hold for both the charge carrier current and the measured total current including the contribution of the displacement current.

The paper is organized as follows. The stochastic approach to describe the $p$-$n$ junction is presented in Sec.~\ref{Sec-th} where we introduce the diffusion-reaction stochastic partial differential equations for the electron and hole densities, as well as the currents.  Under stationary conditions, mean-field equations are deduced from the stochastic ones.  Their dimensionless form is given and the characteristic lengths are identified.  Section~\ref{Sec-num} is devoted to the numerical simulations of the stochastic process, giving the mean density profiles, the mean electric potential, and the current-voltage characteristic curves.  In Sec.~\ref{Sec-FT}, the validity of the fluctuation theorem is established for the charge carrier current and the total current. The conclusion and perspectives are given in Sec.~\ref{Sec-concl}.  Appendices~\ref{AppA} and~\ref{AppB} describe the Markov jump process and the stochastic Langevin equations obtained by spatial discretization of the stochastic partial differential equations for the purpose of numerical simulations.

\section{Stochastic description of the diode}
\label{Sec-th}

\subsection{The $p$-$n$ junction}

\begin{figure*}
\begin{minipage}[t]{0.3\hsize}
\resizebox{1.0\hsize}{!}{\includegraphics{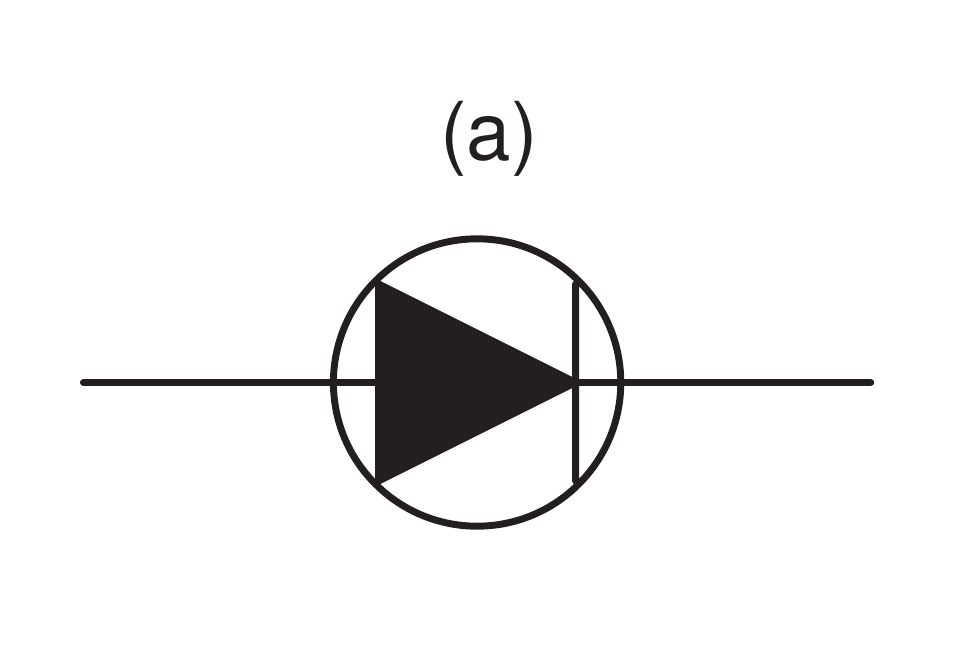}}
\end{minipage}
\begin{minipage}[t]{0.5\hsize}
\resizebox{1.0\hsize}{!}{\includegraphics{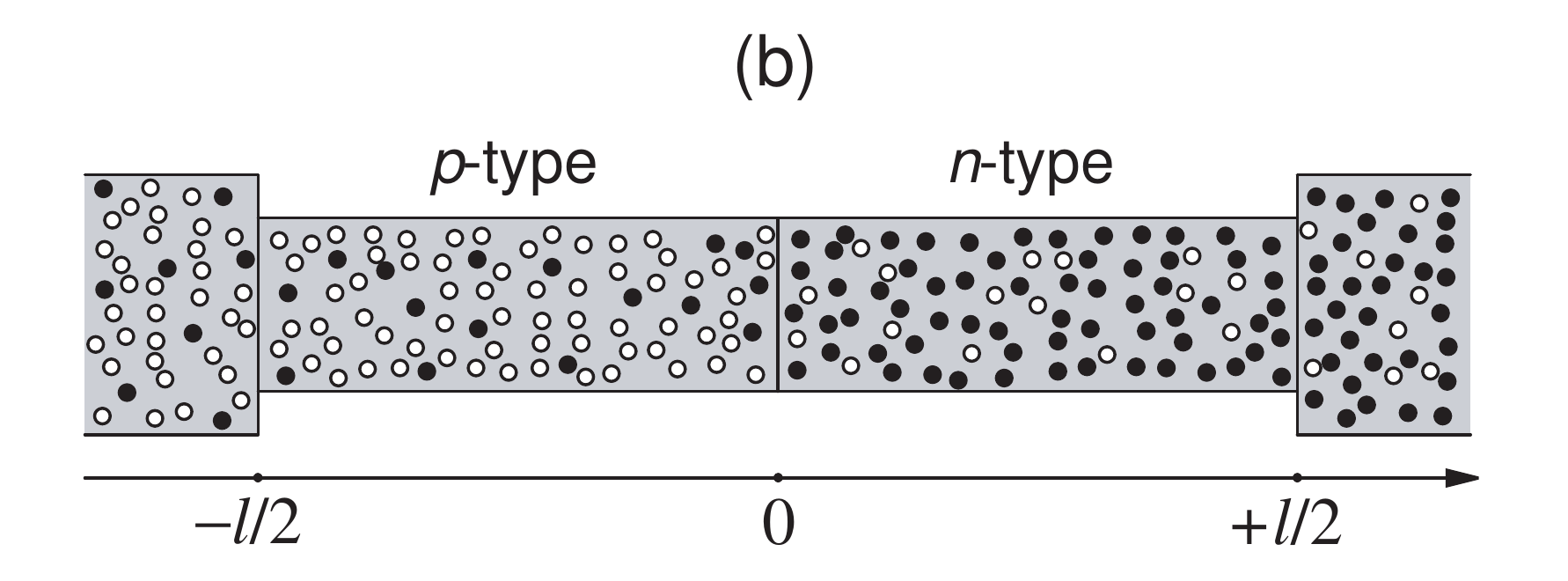}}
\end{minipage}
\caption{Schematic representation of diode (left) and $p$-$n$ junction (right). In the $p$-$n$ junction, the black dots represent electrons and the white ones represent holes.}
\label{fig1}
\end{figure*}

The diode consists of a $p$-$n$ junction, which is composed of a $p$-type semiconductor in contact with a $n$-type semiconductor, as shown in Fig.~\ref{fig1} \cite{AM76}. There are four kinds of charged particles across the $p$-$n$ junction: mobile electrons ${\rm e}^-$, mobile holes ${\rm h}^{+}$, fixed negative-charged acceptors, and fixed positive-charged donors. Inside $p$-type semiconductors, the majority of electric carriers are holes and the minority carriers are electrons. The situation is opposite in $n$-type semiconductors. We can consider the $p$-$n$ junction as a three-dimensional rod of length $l$ with its coordinate $x$ extending from $-l/2$ to $+l/2$, and of section area $\Sigma$ in the transverse $y$- and $z$-directions. The position is denoted ${\bf r}=(x,y,z)$.  The acceptor density $a({\bf r})$ and the donor density $d({\bf r})$ are uniform respectively in the $p$- and $n$-sides.  They can thus be expressed as
\be\label{dfn-a-d}
a({\bf r})=a\, \theta(-x) , \qquad d({\bf r})=d\, \theta(x) ,
\ee
where $\theta(x)$ is Heaviside's step function, which is defined such that $\theta(x)=0$ if $x<0$ and $\theta(x)=1$ if $x>0$. The charge density is given by
\be
\rho = e (p-n+d-a)
\label{rho}
\ee
in terms of the elementary electric charge $e=\vert e\vert$, the hole density $p$, the electron density $n$, the donor density $d$, and the acceptor density $a$.  The charge density determines the electric potential $\phi$ by Gauss's law and the Poisson equation \cite{J99}.  According to electroneutrality, the inhomogeneity~(\ref{dfn-a-d}) of the acceptor and donor densities is thus responsible for the global asymmetric distribution of mobile electrons and holes across the junction, leading to current rectification by the diode.

In doped semiconductors, electrons and holes are generated by the reaction:
\be
\emptyset \ \underset{k_-}{\stackrel{k_+}{\rightleftharpoons}} \ {\rm e}^{-} + {\rm h}^{+} \, ,
\label{reaction}
\ee
where $k_{+}$ and $k_{-}$ are respectively the electron-hole generation and recombination rate constants. 

If the semiconductor was at equilibrium, the electron and hole densities would obey the condition
\be
n_{\rm eq}\, p_{\rm eq} = \nu^2 \, , \qquad\mbox{where} \quad \nu=\sqrt{\frac{k_+}{k_-}}
\label{nu}
\ee
is the intrinsic carrier density.  At equilibrium, the electron and hole densities would be moreover related to the electric potential by
\be
n_{\rm eq}({\bf r})\sim {\rm e}^{+\beta e \phi_{\rm eq}({\bf r})} \qquad\mbox{and}\qquad \, p_{\rm eq}({\bf r}) \sim {\rm e}^{-\beta e \phi_{\rm eq}({\bf r})} \, , 
\label{eq-dens}
\ee
where $\beta\equiv (k_{\rm B} T)^{-1}$ is the inverse temperature.

The diode is in contact on its left- and right-hand sides with reservoirs at different potentials and densities for electrons and holes:
\bea
&&\phi(x=-l/2)=\phi_{\rm L}, \ n(x=-l/2)=n_{\rm L}, \ p(x=-l/2)=p_{\rm L}, \quad \ \label{bc-L}\\
&&\phi(x=+l/2)=\phi_{\rm R}, \ n(x=+l/2)=n_{\rm R}, \ p(x=+l/2)=p_{\rm R}. \quad \ \label{bc-R}
\eea
If the diode is at equilibrium, Eq.~(\ref{eq-dens}) holds and the potential difference is related to the densities at the reservoirs according to Nernst potential
\be
\left(\phi_{\rm L}-\phi_{\rm R}\right)_{\rm eq} = \frac{1}{\beta e} \ln\frac{n_{\rm L}}{n_{\rm R}} = 
\frac{1}{\beta e} \ln\frac{p_{\rm R}}{p_{\rm L}} .
\label{Nernst}
\ee
The diode is driven out of equilibrium if the applied voltage difference with respect to Nernst potential,
\be
V = \phi_{\rm L}-\phi_{\rm R} -\frac{1}{\beta e} \ln\frac{n_{\rm L}}{n_{\rm R}} = 
\phi_{\rm L}-\phi_{\rm R} +\frac{1}{\beta e} \ln\frac{p_{\rm L}}{p_{\rm R}} ,
\label{voltage}
\ee
is non vanishing.  As a consequence, there is an electric current flowing in the diode.
The equilibrium state is recovered if the applied voltage is zero, $V=0$.

\subsection{Stochastic diffusion-reaction equations}

Because of thermal fluctuations, electrons and holes undergo erratic motion inside the diode.  Their mobilities are related to their diffusion coefficients, $D_n$ and $D_p$, according to Einstein's relations:
\be
\mu_n = \beta eD_n  \qquad\mbox{and}\qquad \mu_p = \beta eD_p \, .
\ee
Moreover, electrons and holes are generated and they recombine during every reactive events~(\ref{reaction}).  Since these transport and reactive processes are fluctuating, the densities as well as the electric potential obey stochastic partial differential equations.

The balance equations for electrons and holes of respective densities $n$ and $p$ can be expressed as
\bea
&& \partial_t n + \pmb{\nabla}\cdot {\bf j}_n = \sigma_n \, , \label{dndt-c}\\
&& \partial_t p + \pmb{\nabla}\cdot {\bf j}_p = \sigma_p \, , \label{dpdt-c}
\eea
with the current densities:
\bea
&& {\bf j}_n = -\mu_n \, n \, \pmb{\cal E} - D_n \, \pmb{\nabla} n +  \delta{\bf j}_n \, \label{jn}\\
&& {\bf j}_p = +\mu_p \, p \, \pmb{\cal E} - D_p \, \pmb{\nabla} p +  \delta{\bf j}_p \, , \label{jp}
\eea
and equal reaction rate densities:
\be
\sigma_n = \sigma_p = k_+ - k_- \, n \, p + \delta\sigma 
\label{sigma}
\ee
since the same reactive events determine both of them.  The electric field is expressed as
\be
\pmb{\cal E} = -\pmb{\nabla}\phi 
\ee
in terms of the electric potential satisfying the Poisson equation:
\be
\nabla^2\phi = -\frac{\rho}{\epsilon} 
\ee
where $\epsilon$ is the dielectric constant and the electric charge density is given by Eq.~(\ref{rho}).

The contributions of the fluctuations, $\delta{\bf j}_n$, $\delta{\bf j}_p$, and $\delta\sigma$, are Gaussian white noise fields characterized by
\bea
&& \langle \delta{\bf j}_n({\bf r},t) \rangle = \langle \delta{\bf j}_p({\bf r},t) \rangle = 0 , \label{aj}\\
&& \langle \delta{\bf j}_n({\bf r},t)\otimes \delta{\bf j}_n({\bf r}',t') \rangle = \Gamma_{nn}({\bf r},t) \, \delta^3({\bf r}-{\bf r'}) \, \delta(t-t') \, {\boldsymbol{\mathsf 1}} , \quad \label{jn2} \\
&& \langle \delta{\bf j}_p({\bf r},t)\otimes \delta{\bf j}_p({\bf r}',t') \rangle = \Gamma_{pp}({\bf r},t) \, \delta^3({\bf r}-{\bf r'}) \, \delta(t-t') \, {\boldsymbol{\mathsf 1}} , \quad \label{jp2} \\
&& \langle \delta{\bf j}_n({\bf r},t)\otimes \delta{\bf j}_p({\bf r}',t') \rangle = 0 , \label{jnp}\\
&& \langle\delta\sigma({\bf r},t)\rangle = 0 , \label{as}\\
&& \langle\delta\sigma({\bf r},t)\,\delta\sigma({\bf r'},t')\rangle = \Gamma_{\sigma\sigma}({\bf r},t) \, \delta^3({\bf r}-{\bf r'}) \, \delta(t-t') ,  \label{s2}\\
&& \langle \delta\sigma({\bf r},t)\, \delta{\bf j}_n({\bf r}',t') \rangle = \langle \delta\sigma({\bf r},t)\, \delta{\bf j}_p({\bf r}',t') \rangle = 0 , \label{sj}
\eea
where ${\boldsymbol{\mathsf 1}}$ is the $3\times 3$ identity matrix and
\bea
&&\Gamma_{nn}({\bf r},t) \equiv 2\, D_n \, n({\bf r},t) \, , \label{Gnn}\\
&&\Gamma_{pp}({\bf r},t) \equiv 2\, D_p \, p({\bf r},t) \, , \label{Gpp}\\
&&\Gamma_{\sigma\sigma}({\bf r},t) \equiv k_++k_- n({\bf r},t) p({\bf r},t) \label{Gss}
\eea
are the spectral densities of the noises respectively associated with the electron diffusion, hole diffusion, and reaction.

We notice that the current densities can be equivalently written as
\bea
&& {\bf j}_n = - D_n \, {\rm e}^{\beta e\phi} \, \pmb{\nabla}\left( {\rm e}^{-\beta e\phi} \, n \right)+  \delta{\bf j}_n\, , \label{jn-exp}\\
&& {\bf j}_p = - D_p \, {\rm e}^{-\beta e\phi} \, \pmb{\nabla}\left( {\rm e}^{\beta e\phi} \, p \right) +  \delta{\bf j}_p \, . \label{jp-exp}
\eea

 In general, the parameters $D_p$, $D_n$, and $k_{\pm}$ depend on space in an inhomogeneous medium.  For simplicity, we shall assume that they are uniform in each one of the $p$- and $n$-sides.

\subsection{The currents}

The electric charge is locally conserved because the continuity equation
\be
\partial_t\rho + \pmb{\nabla}\cdot{\bf i} = 0 
\ee
holds for the charge density~(\ref{rho}) and the current density ${\bf i}=e({\bf j}_p-{\bf j}_n)$.  

The current intensity, or shortly the current, is defined as the surface integral
\be
{\cal I}\equiv \int{\bf i}\cdot d\pmb{\Sigma}
\label{current}
\ee 
over the section area $\Sigma$ of the diode.  

The experimentally measured electric current is given by adding the contribution of the displacement current to the particle current
\be
\tilde{\cal I}\equiv \int\left({\bf i}+\epsilon\,\partial_t\,\pmb{\cal E}\right)\cdot d\pmb{\Sigma}\, ,
\label{tot-current}
\ee
which defines the total current \cite{J99,BB00}.

In steady states, the mean values of the current~(\ref{current}) and total current~(\ref{tot-current}) are equal 
\be
I=\langle{\cal I}\rangle=\langle\tilde{\cal I}\rangle
\label{mean-current}
\ee
since the mean value of the displacement current vanishes under stationary conditions.  However, the displacement current should be included to describe the fluctuations of the measured electric current.  In particular, the random number of charges crossing the section area $\Sigma$ during the time interval $[0,t]$
\be
C \equiv \frac{1}{e} \int_0^t {\cal I}(t') \, dt'
\label{C-current}
\ee
will in general differ from the measured quantity
\be
\tilde C \equiv \frac{1}{e} \int_0^t \tilde{\cal I}(t') \, dt' \, ,
\label{C-tot-current}
\ee
although their mean values are equal.  The stochastic processes in the diode can be characterized by the probability distributions $P_t(C)$ and $P_t(\tilde C)$, which are investigated here below.

\subsection{Mean-field equations under stationary conditions}

By averaging the balance equations~(\ref{dndt-c})-(\ref{dpdt-c}) and the expressions~(\ref{jn})-(\ref{jp}) over the noises using Eqs.~(\ref{aj}) and~(\ref{as}),  we can obtain mean-field equations for the stationary mean profiles of the densities and the current densities in the $x$-direction.  Together with Gauss's law and the Poisson equation for the electric field and potential, these mean-field equations are given by the following coupled ordinary differential equations (ODEs):
\bea
&& \frac{dn(x)}{dx}=-\frac{j_n(x)}{D_n}-\beta e n(x){\cal E}(x) , \\
&& \frac{dp(x)}{dx}=-\frac{j_p(x)}{D_p}+\beta e p(x){\cal E}(x) , \\
&& \frac{dj_n(x)}{dx}=k_{+}-k_{-}n(x)p(x) , \label{djndx}\\ 
&& \frac{dj_p(x)}{dx}=k_{+}-k_{-}n(x)p(x) ,  \label{djpdx}\\
&& \frac{d{\cal E}(x)}{dx}=\frac{e}{\epsilon}\left[p(x)-n(x)+d(x)-a(x)\right] , \label{dEdx}\\
&& \frac{d\phi(x)}{dx}=-{\cal E}(x),
\eea
with the aforementioned boundary conditions~(\ref{bc-L}) and~(\ref{bc-R}).  We notice that the first four ODEs are nonlinear.  In this set of ODEs, the net current density
\be
j = j_p(x)-j_n(x)
\ee 
is a constant of integration, as a consequence of electric charge conservation.  Moreover, the electric potential does not appear before the last equation, which is thus decoupled from the other ones.  Accordingly, the set can be reduced to the four ODEs for $n(x)$, $p(x)$, $j_n(x)$, and ${\cal E}(x)$. 

\subsection{Dimensionless form of the ODEs}

A rescaling of the variables is needed in order to obtain a dimensionless form of the ODEs.  With this aim, the intrinsic carrier density $\nu$ introduced in Eq.~(\ref{nu}) is used to define the dimensionless densities
\be
n_*\equiv n/\nu \qquad\mbox{and} \qquad p_*\equiv p/\nu \, .
\ee
The dimensionless time is defined as
\be
t_*\equiv t/\tau\, , \qquad\mbox{where} \qquad \tau = \frac{1}{k_-\nu} = \frac{1}{\sqrt{k_+k_-}}
\label{t-dimless}
\ee
is the intrinsic carrier lifetime.  Supposing that the diffusion coefficients are equal, $D_n=D_p\equiv D$, the position is rescaled as
\be
x_*\equiv x/\ell \, , \qquad\mbox{where} \qquad \ell\equiv \sqrt{D\tau} = \sqrt{\frac{D}{\sqrt{k_+k_-}}}
\label{x-dimless}
\ee
is the intrinsic carrier diffusion length before recombination.  As a consequence of these definitions, the dimensionless current densities are given by
\be
j_{n*}\equiv \frac{j_n}{\nu\ell/\tau}  \qquad\mbox{and} \qquad j_{p*}\equiv\frac{j_p}{\nu\ell/\tau} \, ,
\ee
the electric field and potential by
\be
{\cal E}_* \equiv \ell \beta e{\cal E}  \qquad\mbox{and} \qquad \phi_*\equiv \beta e \phi \, ,
\ee
and the charge numbers by $C_*\equiv C/(\ell^3\nu)$ and $\tilde C_*\equiv \tilde C/(\ell^3\nu)$.
The dimensionless boundary values $n_{\rm L*}$, $p_{\rm L*}$, $\phi_{\rm L*}$, $n_{\rm R*}$, $p_{\rm R*}$, $\phi_{\rm R*}$, applied voltage $V_*$, and diode length $l_*$ are defined in a similar way.

Accordingly, the set of ODEs takes the following dimensionless form:
\bea
&& \frac{dn_*}{dx_*}=-j_{n*}-n_*{\cal E}_* \text{,} \label{eq_ode1} \\
&& \frac{dp_*}{dx_*}=-j_{*}-j_{n*}+p_*{\cal E}_* \text{,} \label{eq_ode2} \\
&& \frac{dj_{n*}}{dx_*}=1-n_*p_* \text{,} \label{eq_ode3} \\ 
&& \frac{d{\cal E}_*}{dx_*}=\alpha\left( p_*-n_*+d_*-a_*\right) \text{,} \label{eq_ode4} \\
&& \frac{d\phi_*}{dx_*}=-{\cal E}_* \text{,} \label{eq_ode5}
\eea
with $j_{p*}=j_{*}+j_{n*}$.  These ODEs only depend on the unique dimensionless parameter 
\be
\alpha\equiv \left(\frac{\ell}{\lambda}\right)^2 = \epsilon^{-1} \beta e^2 \nu D\tau \, ,
\label{alpha}
\ee
where $\ell$ is the intrinsic diffusion length introduced in Eq.~(\ref{x-dimless}) and
\be
\lambda\equiv\sqrt{\frac{\epsilon}{\beta e^2\nu }}
\label{Debye}
\ee
is the intrinsic Debye screening length.

With the boundary conditions at both ends of the diode, the above ODE system constitutes a typical two-point value problem that cannot be analytically solved because of its nonlinearity.  
The discontinuity of $a_*(x_*)$ and $d_*(x_*)$ at the junction makes it hard to solve, so that we use the following continuous alternatives as approximations
\be
a_*(x_*)=\frac{a_*}{1+\exp(x_*/\delta)} \, , \qquad d_*(x_*)=\frac{d_*}{1+\exp(-x_*/\delta)}
\ee
the width $\delta$ being sufficiently small.  In the following, we use the value $\delta=0.01$. 

We note that the dimensionless spectral densities~(\ref{Gnn})-(\ref{Gss}) are given by
\bea
&&\Gamma_{nn*}({\bf r}_*,t_*) = \frac{2}{\ell^3\nu} \, n_* \, , \\
&&\Gamma_{pp*}({\bf r}_*,t_*) = \frac{2}{\ell^3\nu} \, p_* \, , \\
&&\Gamma_{\sigma\sigma*}({\bf r}_*,t_*)= \frac{1}{\ell^3\nu} \, (1+n_* p_*) \, , 
\eea
showing that the noise amplitudes increase if the number $\ell^3\nu$ of intrinsic carriers in a volume of size given by the diffusion length $\ell$ decreases.

\subsection{Characteristic lengths in a homogeneous medium}

In order to determine the characteristic lengths of the density profiles in homogeneous semiconductors of $n$- or $p$-type, Eqs.~(\ref{eq_ode1})-(\ref{eq_ode4}) are linearized around a uniform stationary solution satisfying $n_*p_*=1$ and $p_*-n_*=a_*-d_*$ and given by
\bea
&& n_* = - \frac{a_*-d_*}{2} + \sqrt{\left(\frac{a_*-d_*}{2}\right)^2 + 1} \, , \\
&& p_* = + \frac{a_*-d_*}{2} + \sqrt{\left(\frac{a_*-d_*}{2}\right)^2 + 1} \, , \\
&& j_{n*}= - \frac{n_* j_*}{n_*+p_*} \, , \\
&& j_{p*}= + \frac{p_* j_*}{n_*+p_*} \, , \\
&& {\cal E}_* = \frac{j_*}{n_*+p_*} \, ,
\eea
showing that the electric field and the net current density $j_*$ are proportional to each other and equal to zero at equilibrium.

Small perturbations of this uniform stationary solution obey a set of linear ODEs that have the following matrix form obtained by taking variation of Eqs.~(\ref{eq_ode1})-(\ref{eq_ode4}) up to first order:
\be
\frac{d}{dx_*}\begin{pmatrix} 
\delta n_* \\ 
\delta p_* \\ 
\delta j_{n*} \\ 
\delta {\cal E}_* 
\end{pmatrix}=
\begin{pmatrix}
-{\cal E}_* & 0 & -1 & -n_* \\
0 & {\cal E}_* & -1 & p_* \\
-p_* & -n_* & 0 & 0 \\
-\alpha & \alpha & 0 & 0 
\end{pmatrix}
\begin{pmatrix} 
\delta n_* \\ 
\delta p_* \\ 
\delta j_{n*} \\ 
\delta {\cal E}_* 
\end{pmatrix} .
\label{eq_matrix} 
\ee
Supposing that the perturbations are given by a linear superposition of exponential functions $\delta n_*\sim \exp(\kappa_* x_*)$ and $\delta p_*\sim \exp(\kappa_* x_*)$, we find that
\bea
&& \kappa_*^4 - \left[ (\alpha+1)(n_*+p_*) +{\cal E}_*^2\right] \kappa_*^2 \nonumber\\
&&+ (\alpha-1)(n_*-p_*){\cal E}_*\kappa_* +\alpha (n_*+p_*)^2 = 0 \, .
\eea
Consequently, the four characteristic lengths of the medium are given close to equilibrium by
\bea
&&{\cal L} = \frac{\ell}{\kappa_*} = \pm \sqrt{\frac{D\tau\nu}{n+p}} \, \left[ 1 \pm O(j_*)\right] , \label{diff-length}\\
&&{\cal L}' = \frac{\ell}{\kappa_*'} = \pm \sqrt{\frac{\epsilon}{\beta e^2(n+p)}} \, \left[ 1 \pm O(j_*)\right] . \label{Debye-length}
\eea
These characteristic lengths manifest themselves, especially, at the junction between the semiconductors of $n$- and $p$-types.

\subsection{The Shockley regime}

The stochastic process of charge transfers between the two sides of the junction can be approximately described by the following master equation
\bea
\frac{d}{dt}P(C,t)&=&W^{(+)}P(C-1,t)+W^{(-)}P(C+1,t)\nonumber\\
&&-(W^{(+)}+W^{(-)})P(C,t) ,
\label{master-eq}
\eea
ruling the probability distribution $P(C,t)$ that a number $C$ of charges have been exchanged from the $p$- to the $n$-type side of the junction during the time $t$ \cite{H68,AWBMJ91,AG06}.  $W^{(+)}$ and~$W^{(-)}$ denote the transition rates of charges respectively towards the $n$- and $p$-type sides. In general, both of these rates have a complicated dependence on the applied voltage $V$.

In the case there is a sharp potential barrier at the junction, the transition rate $W^{(-)}$ to descend the barrier becomes independent of the voltage, while the transition rate to climb the barrier can be expressed as $W^{(+)}=W^{(-)}\exp(\beta e V)$ in terms of the Boltzmann factor $\exp(\beta e V)$. Since the mean electric current is given by
\be
I=\lim_{t\to\infty} \frac{e}{t} \, \langle C\rangle_t = e\left(W^{(+)}-W^{(-)}\right),
\label{I-S}
\ee
where $\langle\cdot\rangle_t$ denotes the statistical average with respect to the probability distribution $P(C,t)$, we obtain the Shockley expression for the current-voltage relation \cite{S49,AM76}
\be
I(V)=I_{\rm s}\left[\exp\left(\frac{eV}{k_{\rm B}T}\right)-1\right] ,
\label{eq_Shockley_curve}
\ee
where $I_{\rm s}=eW^{(-)}$ is the residual current of the diode that is independent of the voltage.  However, the Shockley expression only holds in the limit where the majority carriers have a much larger concentration than the minority carriers, as we shall see in the next Sec.~\ref{Sec-num}.

\section{Numerical simulations of the diode model}
\label{Sec-num}

\subsection{Numerical method}

In order to simulate numerically the stochastic process described by Eqs.~(\ref{dndt-c})-(\ref{Gss}), space is discretized into $L$ cells of length $\Delta x=l/L$, section area $\Sigma$, and volume $\Omega=\Sigma\Delta x$, located at the coordinates $x_i=(i-0.5)\Delta x$.  Each cell contains certain numbers of electrons and holes, $N_i=n(x_i)\Omega$ and $P_i=p(x_i)\Omega$.  These numbers change every time a particle jumps between two neighboring cells, between a reservoir and the neighboring cell, or a reactive event occurs generating or recombining an electron-hole pair.  Accordingly, a Markov jump process may be introduced as shown in Appendix~\ref{AppA}, where we give the master equation ruling the time evolution of the probability ${\cal P}({\bf N},{\bf P},t)$ to find the system with the numbers ${\bf N}=(N_i)_{i=1}^L$ of electrons and ${\bf P}=(P_i)_{i=1}^L$ of holes for time $t$.  This Markov jump process can be simulated with Gillespie's algorithm \cite{G76}, which is an exact method for generating the random trajectories of this process.  

A faster simulation method is provided by the corresponding Langevin stochastic process presented in detail in Appendix~\ref{AppB}.  This other process can be deduced from the Markov jump process in the limit where the particle numbers are large enough in the cells.  This method describes the process in terms of stochastic differential equations of Langevin's type given by Eqs.~(\ref{dNdt})-(\ref{xieta2c}) simulating the random time dependence of the particle numbers ${\bf N}(t)$ and ${\bf P}(t)$.  After the discretization of time into equal intervals $\Delta t$, the evolution can be simulated with a recurrence involving independent Gaussian random variables.  In the continuum limit where $\Delta x\to 0$ and $\Delta t\to 0$, the densities of electrons and holes obeying Eqs.~(\ref{dndt-c})-(\ref{Gss}) are recovered through $n(x_{i},t)=N_{i}(t)/\Omega$ and $p(x_{i},t)=P_{i}(t)/\Omega$.  Moreover, the current and total current can be expressed in the framework of Langevin's algorithm as shown in Appendix~\ref{AppB}.

In equilibrium or nonequilibrium stationary states, every mean quantity in some cell can be estimated by the time average
\be
\langle X\rangle=\lim_{T\to\infty}\frac{1}{T}\int_{0}^{T}X(t)\, dt , 
\ee
with $X=N_{i}$, $P_{i}$, $\phi_{i}$, the fluxes $F_i^{(P)}$, $F_i^{(N)}$, or the currents~(\ref{current-discrete}) and (\ref{tot-current-Langevin}).  Due to the electron-hole pair generation and recombination, the mean fluxes $\langle F_i^{(P)}\rangle$ and $\langle F_i^{(N)}\rangle$ vary with the position $x_i$. However, by charge conservation, their difference $\langle F_i^{(P)}\rangle-\langle F_i^{(N)}\rangle$ is independent of the position in the junction, which defines the mean current~(\ref{mean-current}) in the diode. In the following, we study the influence of the different parameters of the system on the profiles of mean electron and hole densities, mean electric potential, and mean current, in order to explore the properties of the junction, especially, the nonlinear response of the current to the applied voltage.  

We consider the boundary conditions
\be
p_{\rm L}=n_{\rm L}+a(-l/2) \, , \qquad n_{\rm R}=p_{\rm R}+d(+l/2) 
\ee
so that the electric field remains uniform in contact with the reservoirs according to Eq.~(\ref{dEdx}).  Furthermore, we suppose
\be
a=d \, , \qquad n_{\rm L}=p_{\rm R} \, , \qquad n_{\rm R}=p_{\rm L} \, ,
\ee
so that the boundary conditions for the particle densities are symmetric under inversion and permutation between electrons and holes, and, moreover, the current densities also remain uniform at the reservoirs by Eqs.~(\ref{djndx}) and~(\ref{djpdx}).  With these assumptions, only the potential difference between the ends of the diode is responsible for inducing particle currents and the diode is characterized by the ratio of majority to minority carrier concentrations:
\be
\frac{c_{\rm major}}{c_{\rm minor}} \equiv \frac{p_{\rm L}}{n_{\rm L}} = \frac{n_{\rm R}}{p_{\rm R}} .
\ee

\subsection{Density profiles and electric potential}

\begin{figure*}
\begin{minipage}[t]{0.33\hsize}
\resizebox{1.0\hsize}{!}{\includegraphics{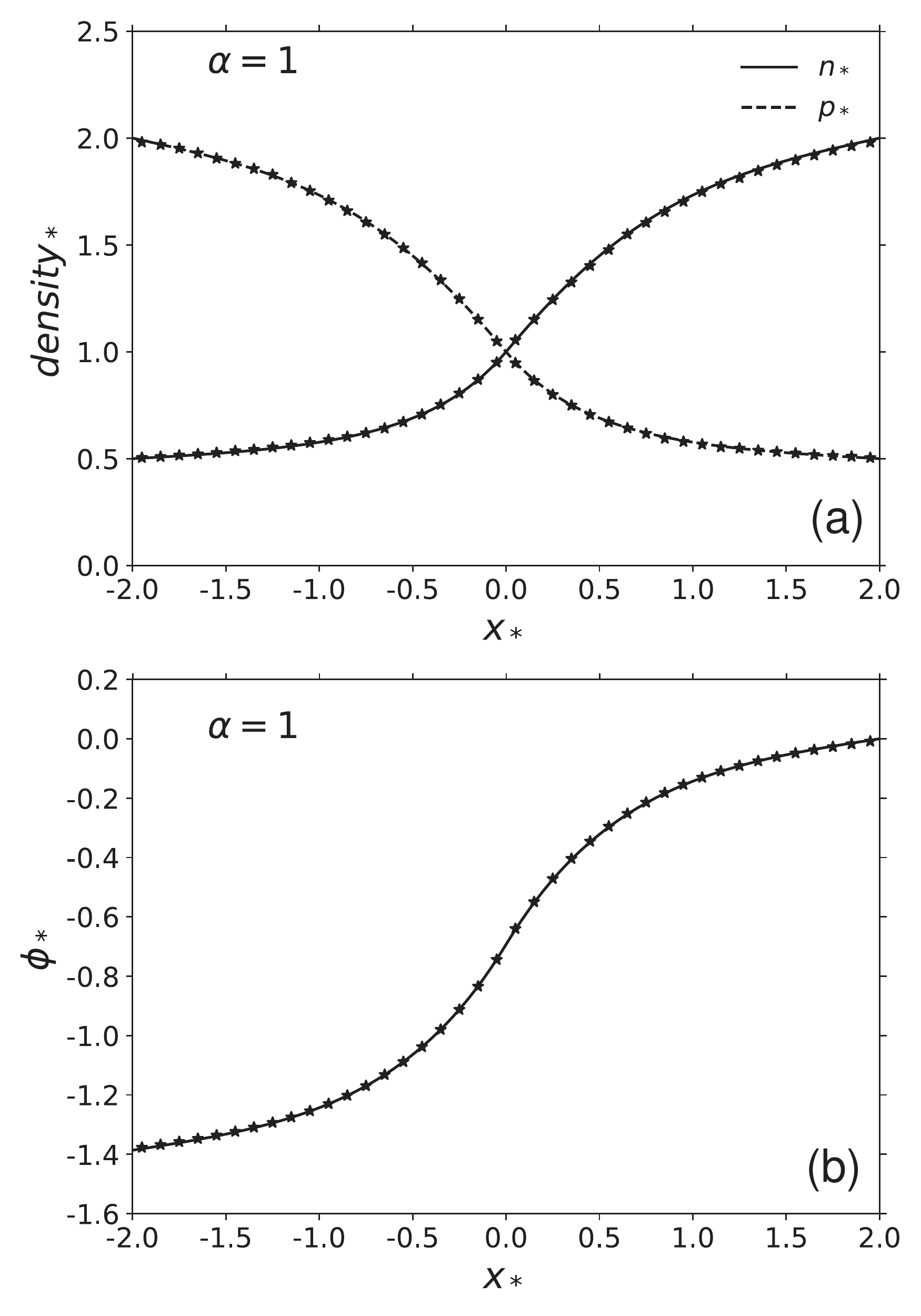}}
\end{minipage}
\begin{minipage}[t]{0.33\hsize}
\resizebox{1.0\hsize}{!}{\includegraphics{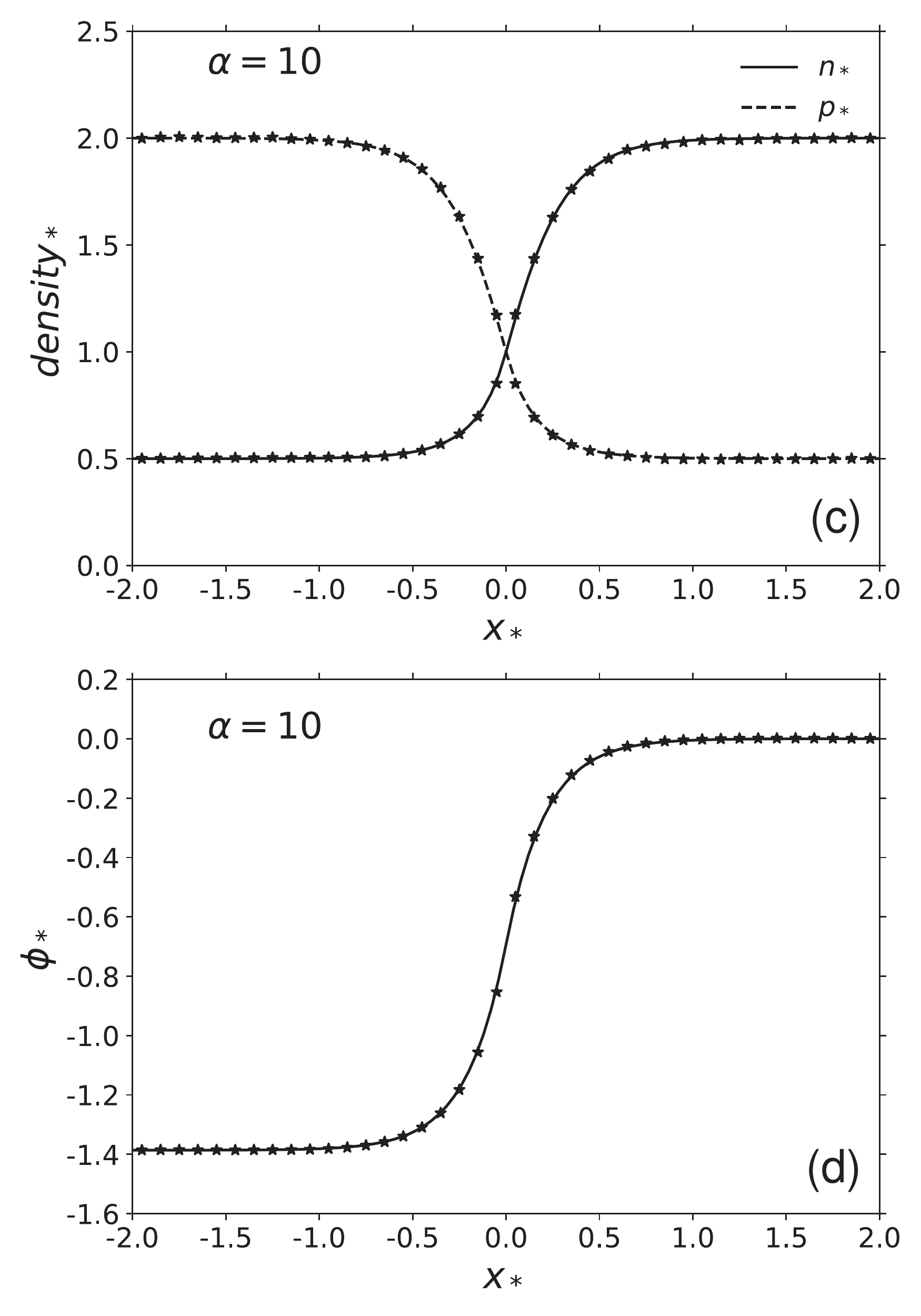}}
\end{minipage}
\begin{minipage}[t]{0.33\hsize}
\resizebox{1.0\hsize}{!}{\includegraphics{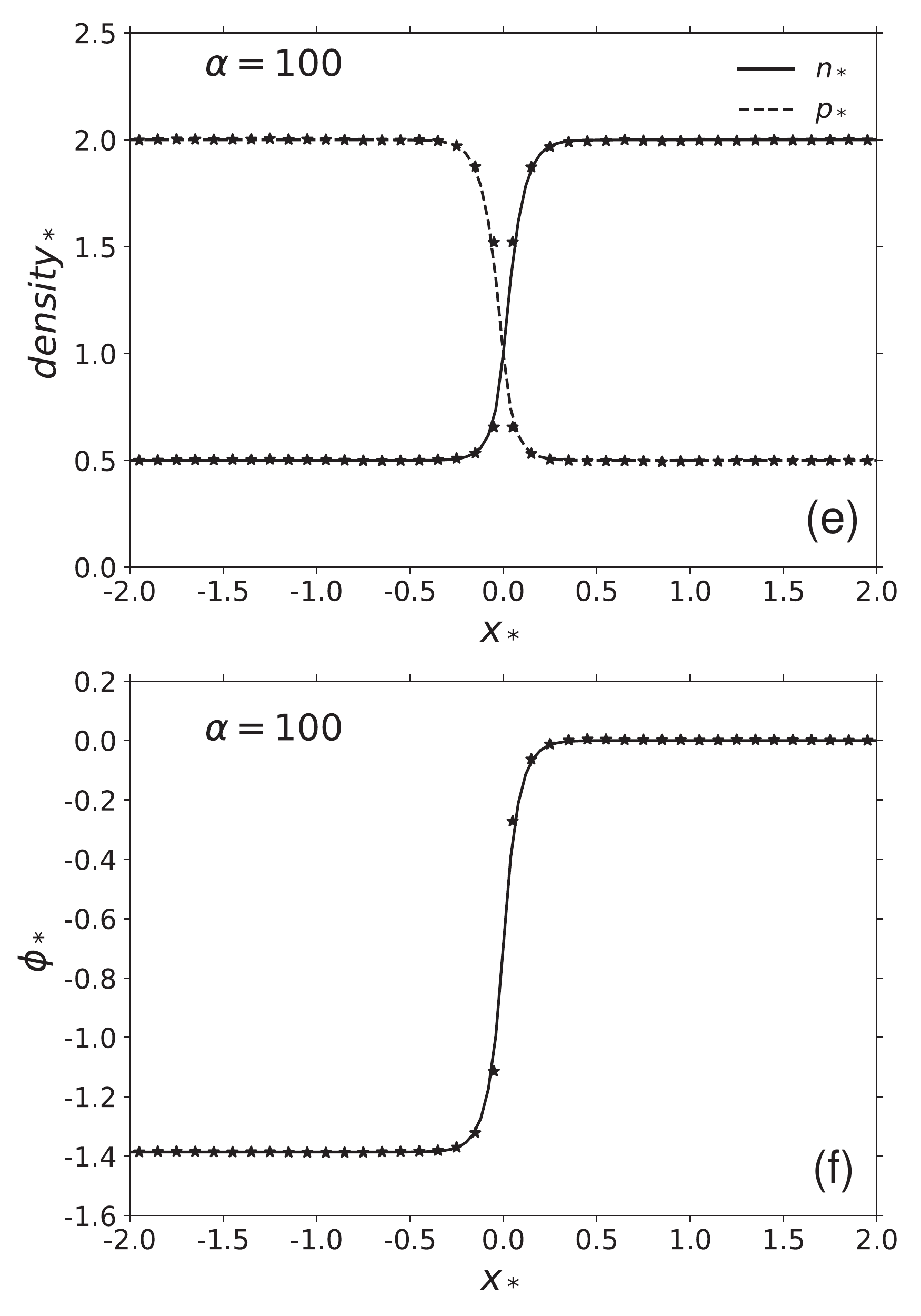}}
\end{minipage}
\caption{The junction for $c_{\rm major}$:$c_{\rm minor}=4$ and different values of the parameter $\alpha$, when the applied voltage is zero $V_*=0$, which implies equilibrium.  Top panels (a), (c), (e): the profiles of the electron density $n_*$ (solid line) and hole density $p_*$ (dashed line). Bottom panels (b), (d), (f): the corresponding profiles of the mean electric potential. The lines depict the profiles obtained by solving the ODEs and the dots by simulating the stochastic process using Langevin's algorithm with $L=40$ cells of volume $\Omega=800$.  The statistics is carried out over $4\times 10^5$ iterates.}
\label{fig2}
\end{figure*}

\begin{figure*}
\begin{minipage}[t]{0.33\hsize}
\resizebox{1.0\hsize}{!}{\includegraphics{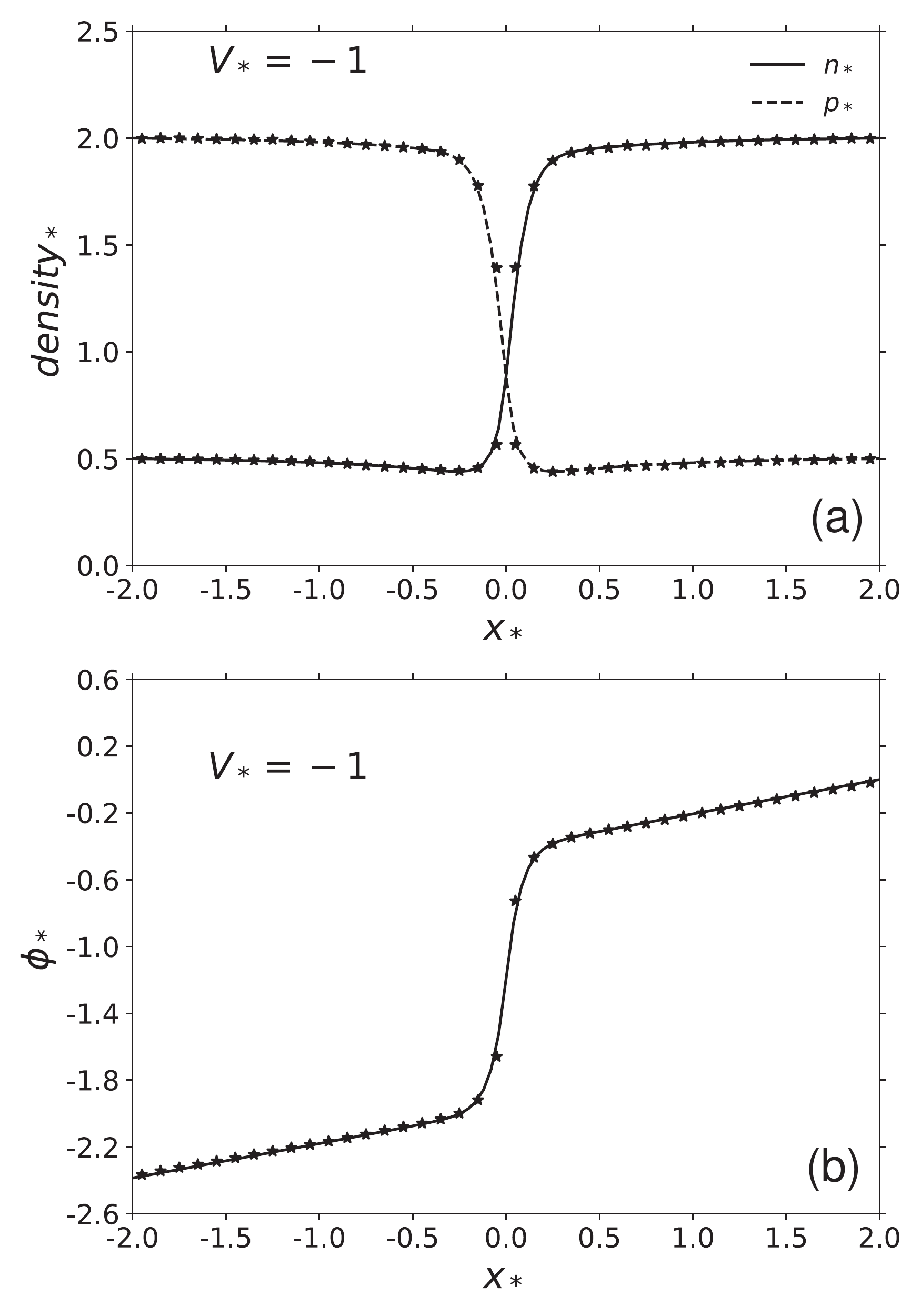}}
\end{minipage}
\begin{minipage}[t]{0.33\hsize}
\resizebox{1.0\hsize}{!}{\includegraphics{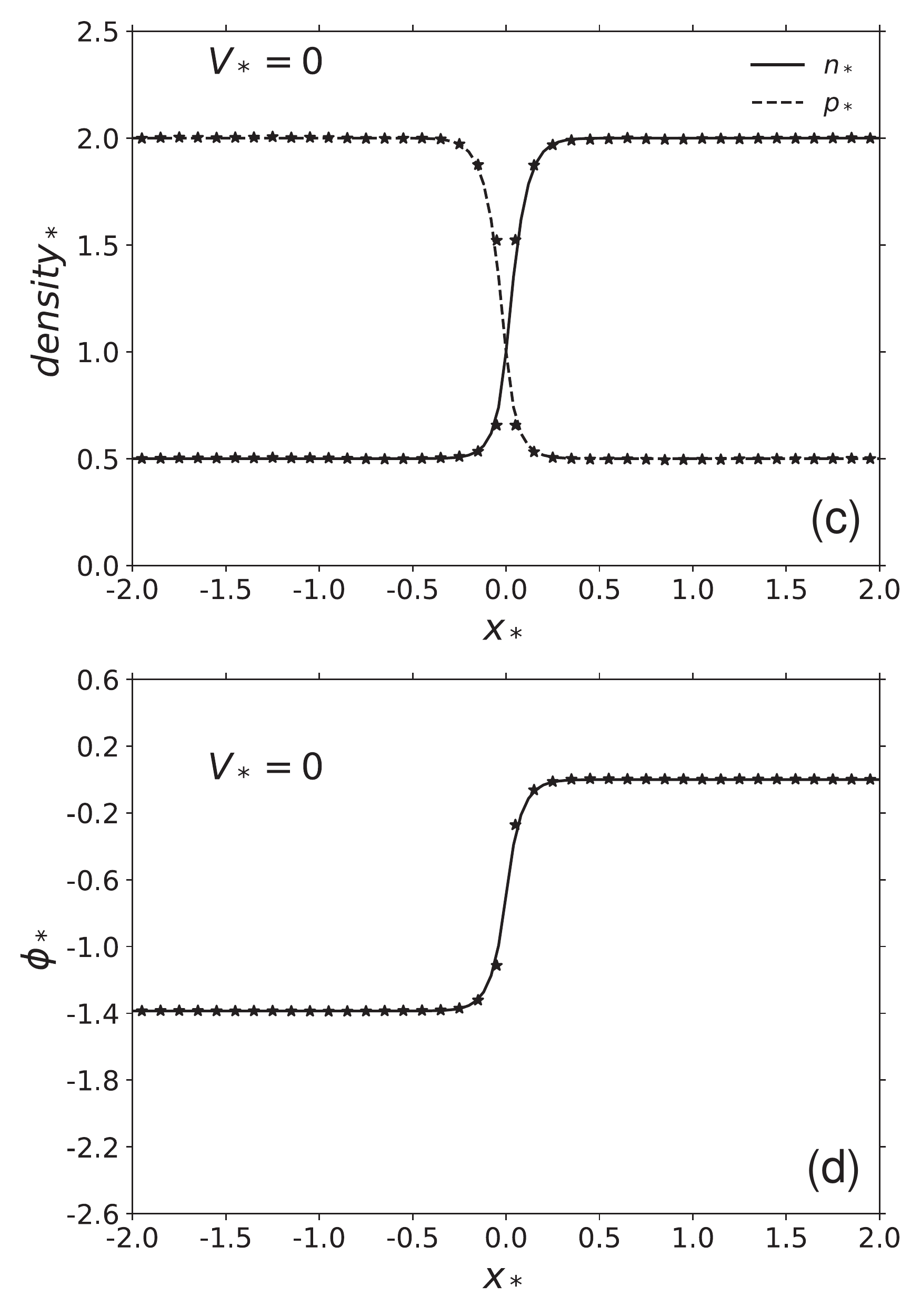}}
\end{minipage}
\begin{minipage}[t]{0.33\hsize}
\resizebox{1.0\hsize}{!}{\includegraphics{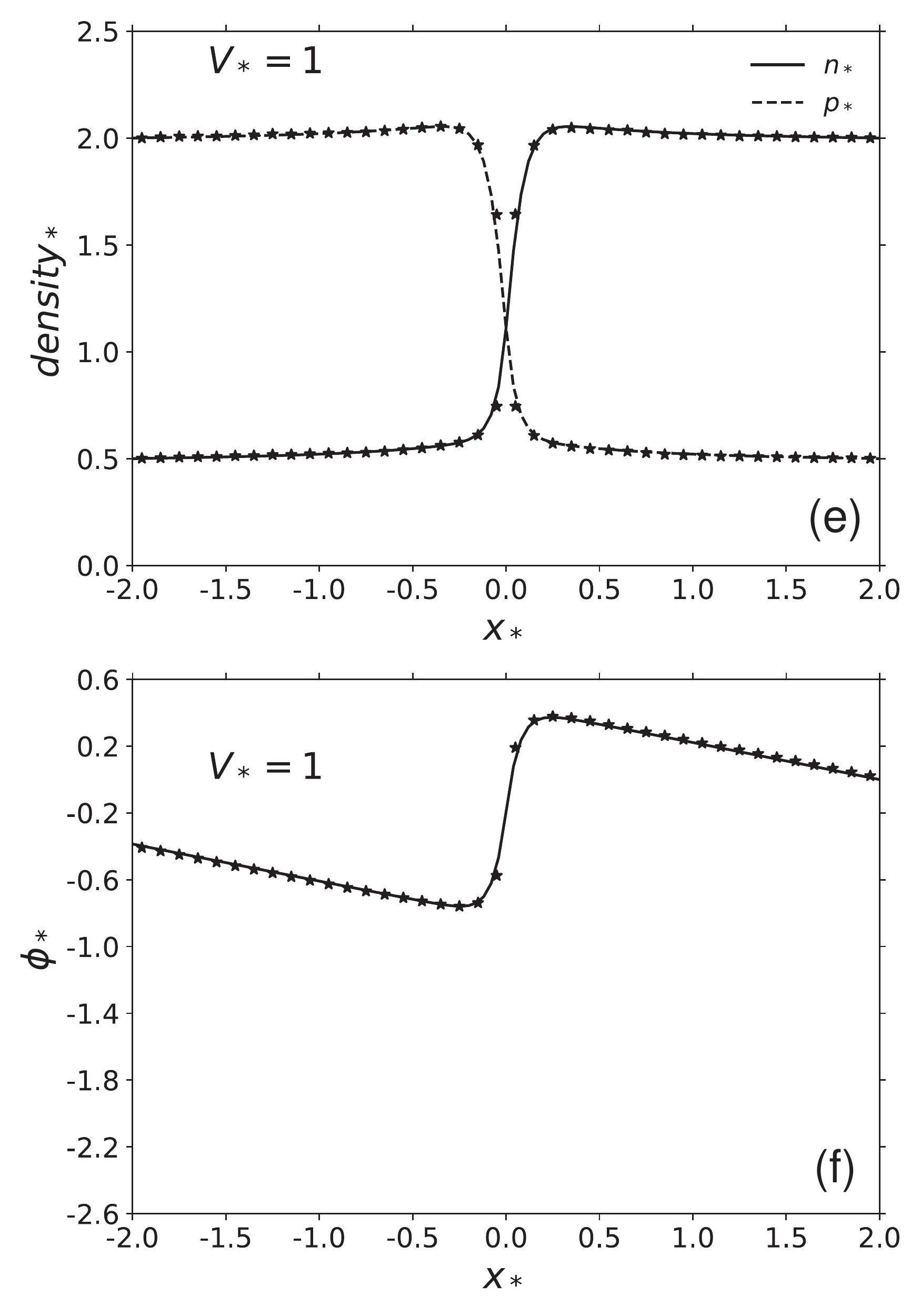}}
\end{minipage}
\caption{The junction for $c_{\rm major}$:$c_{\rm minor}=4$, $\alpha=100$, and different values of the applied voltage $V_*$.  Top panels (a), (c), (e): the profiles of the electron density $n_*$ (solid line) and hole density $p_*$ (dashed line). Bottom panels (b), (d), (f): the corresponding profiles of the mean electric potential. The lines depict the profiles obtained by solving the ODEs and the dots by simulating the stochastic process using Langevin's algorithm with $L=40$ cells of volume $\Omega=800$.  The statistics is carried out over $4\times 10^5$ iterates.}
\label{fig3}
\end{figure*}

Here, we study the effect of the different parameters and boundary conditions on the densities and electric potential across the junction.  In the following, the results of the stochastic simulations are compared in dimensionless form with the solutions of the ODEs~(\ref{eq_ode1})-(\ref{eq_ode5}) giving the mean fields. 

In Fig.~\ref{fig2}, the electron and hole densities, as well as the electric potential, are shown at equilibrium when the applied voltage is equal to zero and for different values of the parameter~(\ref{alpha}).  We observe that the profiles become sharper and sharper as $\alpha$ increases.  In this progression, the Debye screening length~(\ref{Debye}) becomes smaller and smaller with respect to the carrier diffusion length introduced in Eq.~(\ref{x-dimless}).  The bottom panels~(b),~(d), and~(f) of Fig.~\ref{fig2} show that the electric potential is not uniform at equilibrium because of the ceaseless process of electron-hole generation and recombination between the $p$- and $n$-types of semiconductors in contact at the junction \cite{AM76}.  According to Eq.~(\ref{Nernst}), the potential difference calculated from the boundary conditions $p_{{\rm L}*}=n_{{\rm R}*}=2$ and $n_{{\rm L}*}=p_{{\rm R}*}=0.5$ is equal to ($\Delta\phi_*)_{\rm eq}=\ln(2.0/0.5)\simeq 1.38$ when the junction is in equilibrium, as seen in Fig.~\ref{fig2}.  Away from the junction, the densities and the potential become asymptotically uniform because the current and the electric fields are vanishing at equilibrium.  The characteristic length of this approach to uniformity is essentially the Debye screening length given by Eq.~(\ref{Debye-length}).  We see the nice agreement between the results of the mean-field ODEs~(\ref{eq_ode1})-(\ref{eq_ode5}) and those of the simulation. 

Figure \ref{fig3} shows the profiles for different values of the applied voltage $V_*$.  Now, the profiles are deformed by the nonequilibrium constraint of the applied voltage $V_*$.  The slope of the electric potential gives the electric field by Eq.~(\ref{eq_ode5}), which is nonvanishing under the nonequilibrium voltages $V_*=\pm 1$.  In the panels~(a) and~(e) of Fig.~\ref{fig3}, the density profiles are also deformed in their approach towards uniform profiles away from the junction.  Out of equilibrium, the characteristic length~(\ref{diff-length}) of carrier diffusion before recombination manifests itself.  This latter is longer than the Debye screening length~(\ref{Debye-length}) because $\vert{\cal L}/{\cal L}'\vert\simeq \ell/\lambda=\sqrt{\alpha}=10$ in the conditions of Fig.~\ref{fig3}.  As expected~\cite{AM76}, there is an excess of holes on the $p$-type side of the junction under a positive voltage with respect to the situation at equilibrium.  Again, there is a good agreement between the results of the simulation and the mean-field profiles, which brings a strong support to the validity of the stochastic approach.

\subsection{Current-voltage characteristic curve}

The heterogeneous distributions of charge carriers induce the effect of current rectification in the diode.  This rectifying effect is characterized by the nonlinear dependence of the mean current on the voltage.

\begin{figure*}
\begin{minipage}[t]{0.55\hsize}
\resizebox{1.0\hsize}{!}{\includegraphics{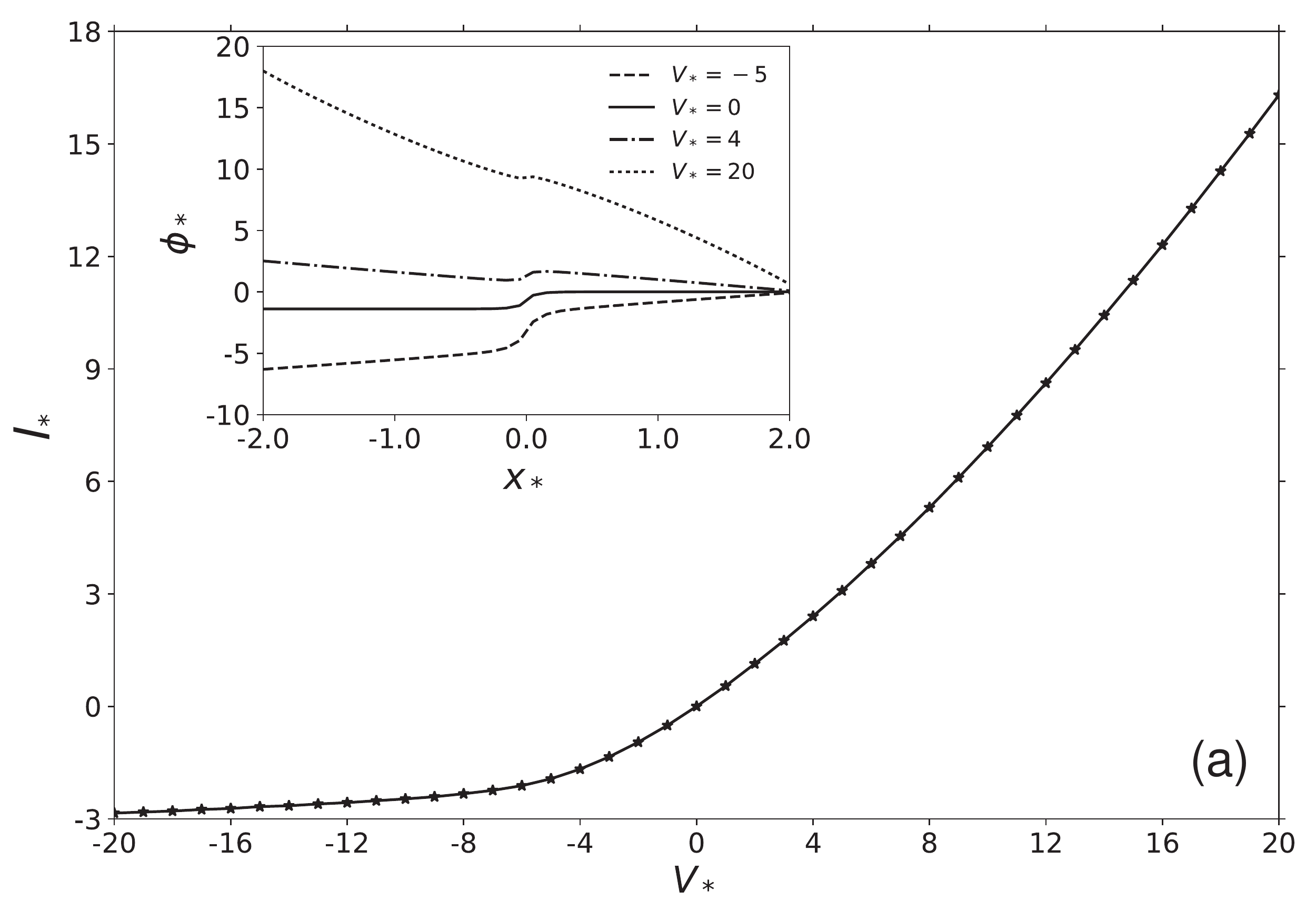}}
\end{minipage}
\begin{minipage}[t]{0.55\hsize}
\resizebox{1.0\hsize}{!}{\includegraphics{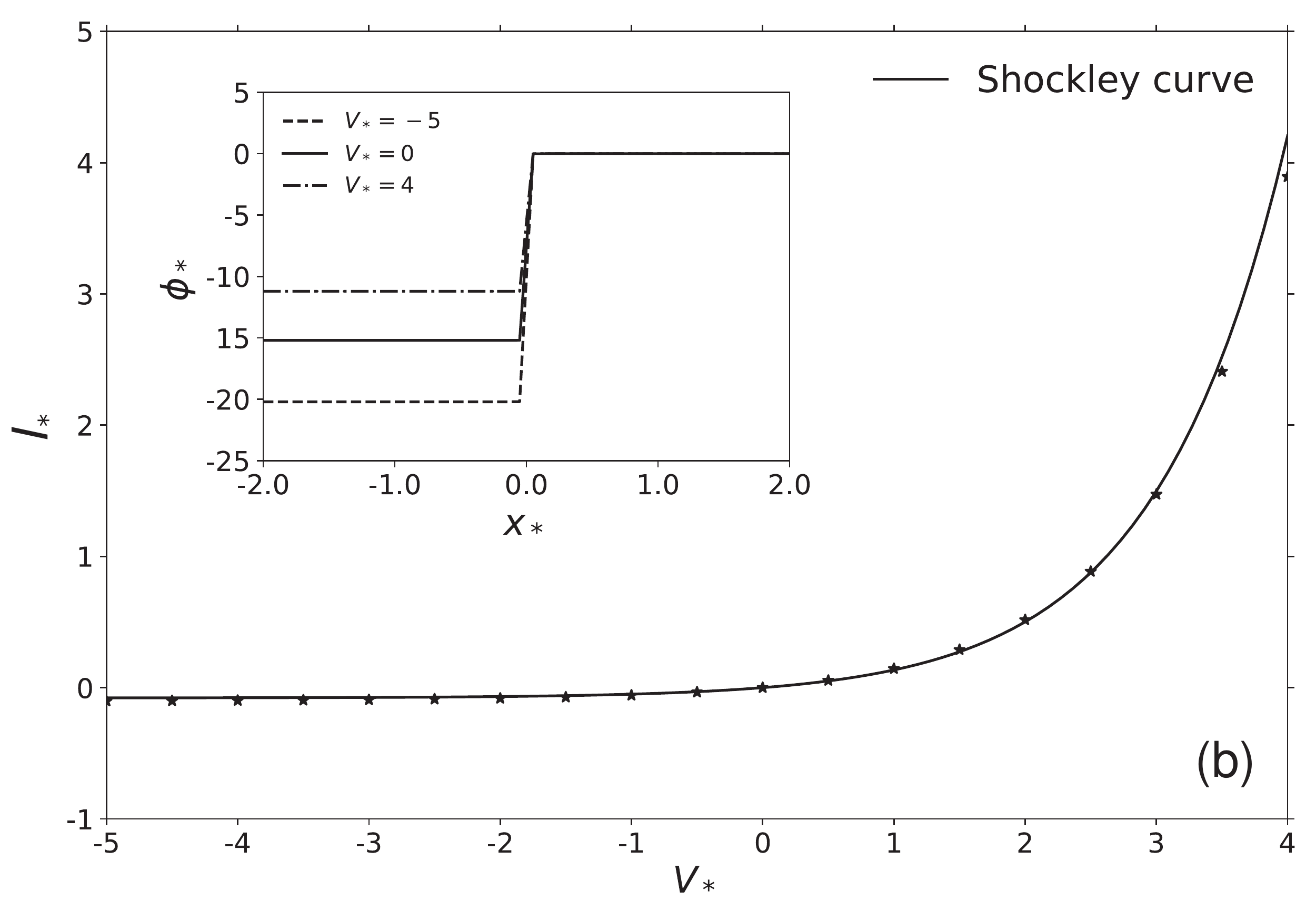}}
\end{minipage}
\caption{The current-voltage characteristic curves of the $p$-$n$ junction with $\alpha=100$: (a) for $c_{\rm major}$:$c_{\rm minor}=4$ and (b) for $c_{\rm major}$:$c_{\rm minor}=4\times 10^6$. $I_*$~denotes the dimensionless mean current and $V_*$ the dimensionless applied voltage. The inset in each panel shows the electric potential profiles under several values of the applied voltage.  The dots show the results of the stochastic simulation.  In panel (a), the line is joining the data points.  In panel (b), the line is the Shockley curve fitted to simulation data.  The simulations are performed using Langevin's algorithm with $L=40$ cells of volume $\Omega=800$ for~(a) and $\Omega=8\times 10^5$ for~(b).  The statistics is carried out over $4\times 10^5$ iterates.}
\label{fig4}
\end{figure*}

We see in Fig.~\ref{fig4}a the current-voltage function of the junction under the same conditions as in Fig.~\ref{fig3}.  As expected, the mean current increases more rapidly for a positive than a negative voltage.  However, the rectification effect is moderate and the $I$-$V$ curve differs from the Shockley function~(\ref{eq_Shockley_curve}) because the barrier of the electric potential at the junction is not sharp enough with respect to the voltage drop across the diode, as seen in the inset of Fig.~\ref{fig4}a.  If the Shockley model~(\ref{eq_Shockley_curve}) would hold, the rectification ratio ${\mathscr R}=\vert I_*(V_*)/I_*(-V_*)\vert$ would be equal to ${\mathscr R}_{\rm S}=\exp(V_*)$.  However, this ratio takes the value ${\mathscr R}\simeq 1.5$ at $V_*=4$, much lower than the expected value ${\mathscr R}_{\rm S}= 54.6$, which confirms that the Shockley model does not apply in the conditions of Fig.~\ref{fig4}a.

In order to reach the Shockley regime, the concentration ratio of majority to minority carriers is increased up to the very large value $c_{\rm major}$:$c_{\rm minor}=4\times 10^6$, so that the potential step takes the value $(\Delta\phi_*)_{\rm eq}=\ln(4\times 10^6)=15.2$ at equilibrium.  Figure~\ref{fig4}b confirms that the $I$-$V$ curve follows the Shockley function~(\ref{eq_Shockley_curve}) in this case with the small value $I_{{\rm s}*}\simeq 0.08$ of the residual current at negative voltage.

After having checked that the stochastic approach describes the expected properties for the mean quantities, we proceed in the next Sec.~\ref{Sec-FT} with the study of fluctuation properties.

\section{Fluctuation theorem for electric currents}
\label{Sec-FT}

\subsection{Current fluctuations}

\begin{figure*}
\begin{minipage}[t]{1.0\hsize}
\resizebox{1.02\hsize}{!}{\includegraphics{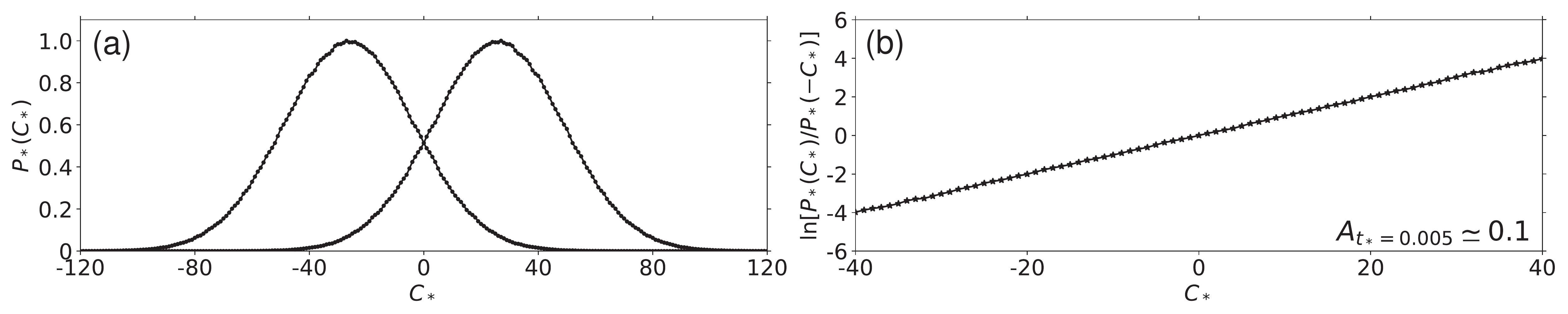}}
\end{minipage}
\begin{minipage}[t]{1\hsize}
\resizebox{1.02\hsize}{!}{\includegraphics{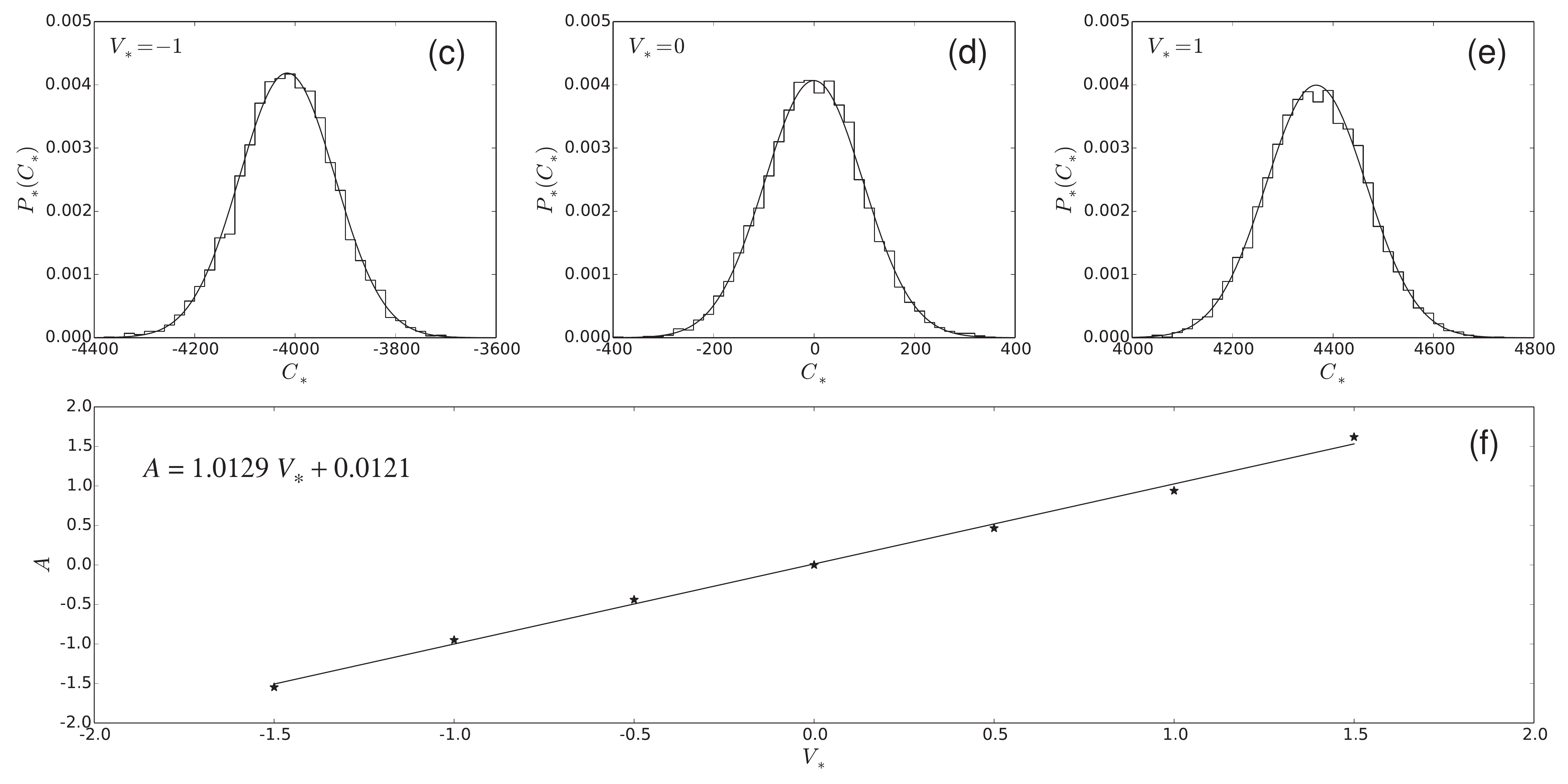}}
\end{minipage}
\caption{Full-counting statistics of the carrier electric current through the section area at the middle of the junction for $c_{\rm major}$:$c_{\rm minor}=4$, $\alpha=100$, and $l_*=4$: (a) The probability distributions $P_*(\pm C_*)$ for time $t_*=0.005$ versus the rescaled charge number $C_*=C/(\ell^3\nu)$ with peak value normalized to one. The statistics is carried out over $10^6$ random trajectories. (b) $\ln[P_*(C_*)/P_*(-C_*)]$ versus $C_*$ showing the linearity with the slope $A_{t*=0.005}\simeq 0.1$ for time $t_*=0.005$. (c)-(d)-(e) Histograms $P_*(C_*)$ of the carrier charge distributions for time $t_*=1$ and different values of the dimensionless applied voltage (c)~$V_*=-1$, (d)~$V_*=0$, and (e)~$V_*=1$. The histograms are obtained with $5\times 10^4$ data and they are fitted to Gaussian distributions~(\ref{eq_gaussian}). (f) The affinities computed with these fitted Gaussian distributions according to Eq.~(\ref{At-formula}) versus the dimensionless applied voltage $V_*$, checking the linear dependence $A\simeq V_*$ with a unit slope up to numerical accuracy (solid line) for time $t_*=1$. 
The simulations are performed using Langevin's algorithm with $L=40$ cells of volume $\Omega=800$.}
\label{fig5}
\end{figure*}

\begin{figure*}
\begin{minipage}[t]{1.0\hsize}
\resizebox{1.02\hsize}{!}{\includegraphics{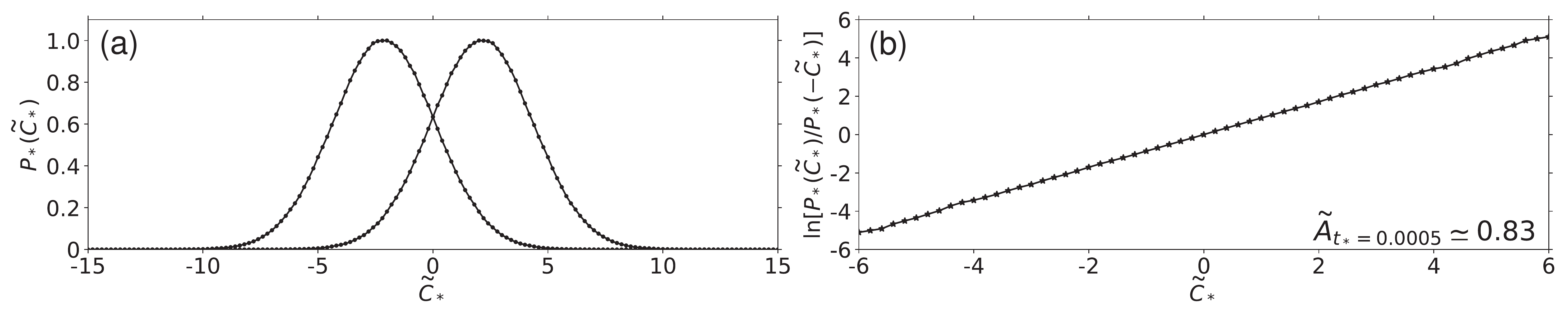}}
\end{minipage}
\begin{minipage}[t]{1\hsize}
\resizebox{1.02\hsize}{!}{\includegraphics{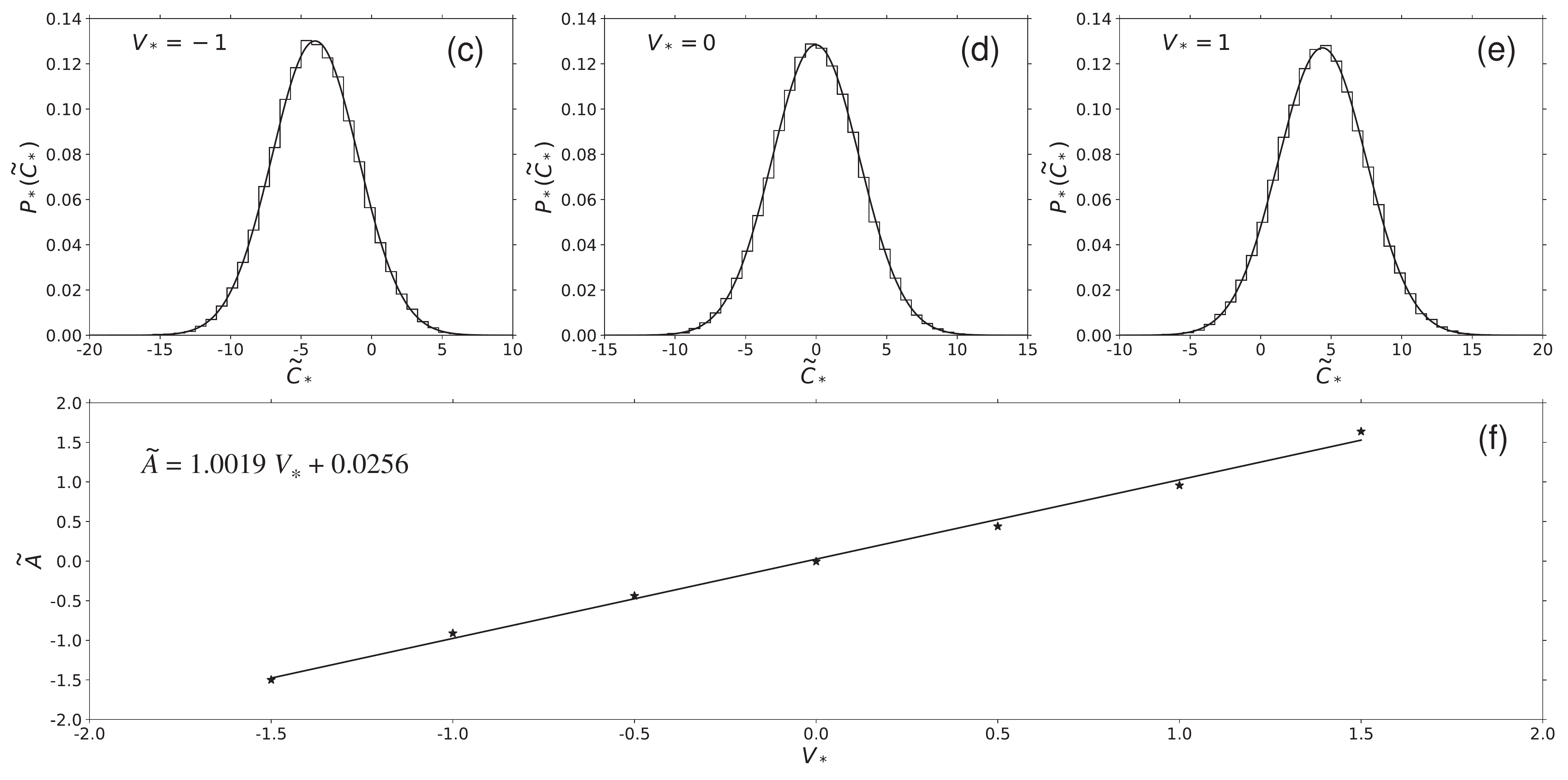}}
\end{minipage}
\caption{Full-counting statistics of the total electric current including the contribution of the displacement current for $c_{\rm major}$:$c_{\rm minor}=4$, $\alpha=100$, and $l_*=4$: (a) The probability distributions $P_*(\pm \tilde C_*)$ for time $t_*=0.0005$ versus the rescaled charge number $\tilde C_*=\tilde C/(\ell^3\nu)$ with peak value normalized to one. The statistics is carried out over $10^6$ random trajectories. (b) $\ln[P_*(\tilde C_*)/P_*(-\tilde C_*)]$ versus $\tilde C_*$ showing the linearity with the slope $\tilde A_{t*=0.0005}\simeq 0.83$ for time $t_*=0.0005$. (c)-(d)-(e) Histograms $P_*(\tilde C_*)$ of the total charge distributions for time $t_*=0.001$ and different values of the dimensionless applied voltage (c)~$V_*=-1$, (d)~$V_*=0$, and (e)~$V_*=1$. The histograms are obtained with $5\times 10^4$ data and they are fitted to Gaussian distributions~(\ref{eq_gaussian}). (f) The affinities computed with these fitted Gaussian distributions according to Eq.~(\ref{At-formula}) versus the dimensionless applied voltage $V_*$, checking the linear dependence $\tilde A\simeq V_*$ with a unit slope up to numerical accuracy (solid line) for time $t_*=0.001$. The simulations are performed using Langevin's algorithm with $L=40$ cells of volume $\Omega=800$.}
\label{fig6}
\end{figure*}

Here, we consider the fluctuations of the electric current~(\ref{current}) in the middle of the junction at the location $x=0$.  This current is composed of electrons and holes moving in either directions and crossing the section area $\Sigma$ at $x=0$ for random times $t_n$.  Accordingly, this instantaneous current can be written as
\be
{\cal I}(t) = \sum_{n=-\infty}^{+\infty} q_n \, \delta(t-t_n) 
\ee
with $q_n=\pm e$ depending on whether the carrier is positively or negatively charged and moves towards $x>0$ or $x<0$.  This random process is ruled by the master equation~(\ref{eq_master_equation}) given in Appendix~\ref{AppA}.  The system is driven out of equilibrium by constraints at its boundaries.  Using the network representation of the master equation put forward by Schnakenberg \cite{S76}, macroscopic affinities can be identified from the cyclic paths of the graph associated with the Markov jump process.  As shown in Subsec.~\ref{AppA4}, the macroscopic affinity corresponding to the transfer of one unit charge from the left to the right reservoir is given by
\be
A=\ln\left[\frac{p_{\rm L}}{p_{\rm R}}\text{e}^{\beta e(\phi_{\rm L}-\phi_{\rm R})}\right]=\beta eV
\label{macro-aff}
\ee
in terms of the applied voltage~(\ref{voltage}), as expected.  Equivalent cyclic paths give the same affinity.  These macroscopic affinities characterize the nonequilibrium driving of the diode.  If the applied voltage is vanishing, the affinities are all equal to zero and we recover the macroscopic equilibrium condition for the diode.

As shown in Ref.~\cite{AG07}, the currents of such a Markov jump process obey a fluctuation theorem.  For the current~(\ref{current}), the fluctuation theorem can be expressed as
\be
\frac{P(C,t)}{P(-C,t)}\simeq_{t\to\infty}\exp(AC) ,
\label{FT-current}
\ee
in terms of the probability $P(C,t)$ that the number~(\ref{C-current}) of charges crossing the section area $\Sigma$ during the time interval~$[0,t]$ is equal to $C$.  At equilibrium if $A=0$, we recover the global detailed balance condition, according to which the probabilities of opposite fluctuations are equal.  Since the Langevin stochastic process described in Appendix~\ref{AppB} is the limit of the Markov jump process for $N_i\gg 1$ and $P_i\gg 1$, the fluctuation theorem also applies to the charge number given by
\be
C = \int_0^t \left[ F_i^{(P)}(t')-F_i^{(N)}(t')\right] dt'
\label{C-current-approx}
\ee
in terms of the fluxes~(\ref{FiN}) and~(\ref{FiP}) at the chosen location~$x_i$.  In the continuum limit, the random variable~(\ref{C-current-approx}) corresponds to the number~(\ref{C-current}) of the stochastic process defined by Eqs.~(\ref{dndt-c})-(\ref{Gss}), so that the fluctuation theorem~(\ref{FT-current}) should apply to the diode.  If the logarithm of the ratio of the probabilities $P(\pm C,t)$ is linear in the random variable $C$, we may consider the affinity
\be
A_t\equiv \frac{1}{C} \ln\frac{P(C,t)}{P(-C,t)}
\label{At}
\ee
for the given time $t$.  The prediction of the fluctuation theorem is that its asymptotic value should be equal to the macroscopic affinity given by the dimensionless applied voltage
\be
\lim_{t\to\infty} A_t = A = \beta e V = V_* \, .
\label{FT-predict}
\ee

In order to test numerically the prediction of the fluctuation theorem, we consider the full counting statistics of the random variable~(\ref{C-current-approx}) as simulated by the Langevin stochastic differential equations~(\ref{dNdt})-(\ref{xieta2c}).  The probability distribution $P_*(C_*,t_*)$ is computed for the rescaled random number $C_*=C/(\ell^3\nu)$ and the dimensionless time introduced in Eq.~(\ref{t-dimless}).  These quantities are shown in Fig.~\ref{fig5}a-b for the short time $t_*=0.005$.  In this case, the probability distributions $P_*(\pm C_*,t_*)$ have a strong overlap so that the affinity~(\ref{At}) can be directly evaluated to be $A_{t_*=0.005}\simeq 0.1$, which is much smaller than the asymptotic value giving the macroscopic affinity $A=A_{\infty}=V_*=1$.

In order to reach the macroscopic affinity, the time interval is increased.  However, the overlap between the probability distributions $P_*(\pm C_*,t_*)$ soon become very small as time increases if $V_*\neq 0$.  Indeed, $P_*(\pm C_*,t_*)$ are distributed around their mean values $\pm\langle C_*\rangle$, which increase linearly with time at rates equal to plus or minus the mean current.  As seen in Fig.~\ref{fig5}c for time $t_*=1$ and voltage $V_*=-1$, the distribution $P_*(C_*,t_*)$ is centered around $\langle C_*\rangle\simeq -4017$ so that its overlap is tiny with the opposite distribution $P_*(-C_*,t_*)$, which is centered around $\langle C_*\rangle\simeq +4017$.  A similar feature holds for the distribution of Fig.~\ref{fig5}e at time $t_*=1$ and voltage $V_*=1$, which has the mean value $\langle C_*\rangle\simeq +4366$.  Therefore, we should use a different method to evaluate the affinity in such circumstances.  In Fig.~\ref{fig5}c-d-e, we observe that the histograms $P_*(C_*,t_*)$ are very close to Gaussian distributions for three different values of the applied voltage.  According to the central limit theorem~\cite{B95}, the probability distribution should indeed take the Gaussian form
\be
P(C,t)\simeq \frac{1}{\sqrt{2\pi \sigma_t^2}}\exp\left[-\frac{(C-\mu_t)^2}{2\sigma_t^2}\right],
 \label{eq_gaussian}
\ee
if there is a large enough number of charge transfers during the time interval $[0,t]$.  In Eq.~(\ref{eq_gaussian}), $\mu_t$ is the mean number of charge transfers and $\sigma_t^2$ the corresponding variance. Introducing the transition rates $W_t^{(\pm)}$ respectively for transfers towards $x>0$ or $x<0$ during the time interval $[0,t]$, the mean flux and diffusivity of charge transfers would be given by
\be
J_t\equiv \frac{\mu_t}{t} = W_t^{(+)}-W_t^{(-)} \, , \qquad D_t\equiv\frac{\sigma_t^2}{2t}=\frac{1}{2}(W_t^{(+)}+W_t^{(-)}) .
\label{Jt-Dt}
\ee
Hence, it is possible to evaluate the affinity with
\be
A_t\simeq \ln\frac{W_t^{(+)}}{W_t^{(-)}}= \ln\frac{2D_t+J_t}{2D_t - J_t} .
\label{At-formula}
\ee
For the particular process ruled by the master equation~(\ref{master-eq}), Eq.~(\ref{At-formula}) indeed gives the corresponding affinity, $A=\ln(W^{(+)}/W^{(-)})=\beta e V$.  

In Fig.~\ref{fig5}, the histograms $P_*(C_*,t_*=1)$ are fitted to Gaussian distributions~(\ref{eq_gaussian}).  The associated mean flux $J_t$, diffusivity $D_t$, and transition rates $W_t^{(\pm)}$ are given by using Eq.~(\ref{Jt-Dt}) and the affinity is thus calculated with Eq.~(\ref{At-formula}).  The values of the so-calculated affinity are plotted in Fig.~\ref{fig5}f as a function of the dimensionless applied voltage $V_*$, showing agreement with the prediction~(\ref{FT-predict}) of the fluctuation theorem for time $t_*=1$.  Therefore, the affinity converges to the macroscopic value~(\ref{macro-aff}), as predicted by the fluctuation theorem~(\ref{FT-current}) for the carrier electric current.

\subsection{Total current fluctuations}

As shown in Appendix~\ref{AppB}, the integral~(\ref{C-tot-current}) of the measured total current during the time interval $[0,t]$ can be expressed as
\be
\tilde C= \frac{1}{L+1} \sum_{j=0}^{L} \int_0^t \left[ F_j^{(P)}(t')-F_j^{(N)}(t')\right] dt'
\label{tilde_C}
\ee
in terms of the fluxes~(\ref{FiN}) and~(\ref{FiP}).  The fluctuation theorem with the macroscopic affinity~(\ref{macro-aff}) also applies to the fluctuations of the random variable~(\ref{tilde_C})
\be
\frac{P(\tilde C,t)}{P(-\tilde C,t)}\simeq_{t\to\infty}\exp(A\tilde C) \, ,
\label{FT-tot-current}
\ee
by extension of the previously considered theorem~\cite{AG07,AG09}.

Figure~\ref{fig6}a-b shows the probability distributions $P(\pm \tilde C,t)$ for time $t_*=0.0005$, as well as the associated affinity~$\tilde A_t$ given by Eq.~(\ref{At}) with $C$ replaced by $\tilde C$.  Remarkably, for this short time, this affinity is already close to the asymptotic value given by the macroscopic value~(\ref{macro-aff}).  In Fig.~\ref{fig6}c-d-e, the histograms are fitted to Gaussian distributions to compute the affinity $\tilde A_t$ for time $t_*=0.001$ and different values of the applied voltage.  In Fig.~\ref{fig6}f, the affinity is plotted versus the applied voltage, showing that they are equal up to numerical accuracy, already for the short time $t_*=0.001$.  Therefore, the fluctuation theorem~(\ref{FT-tot-current}) for the total electric charge is confirmed by these results.  Furthermore, the convergence to the macroscopic affinity is observed to be much faster for the measured total electric charge fluctuations than for the carrier electric charge fluctuations.

\subsection{Thermodynamic entropy production}

An implication of the fluctuation theorem is that the thermodynamic entropy production is always non-negative.  Indeed, the entropy production can be expressed as \cite{G13}
\be
\frac{1}{k_{\rm B}}\frac{d_{\rm i}S}{dt}=\lim_{t\to\infty}\frac{1}{2t} \int d\tilde C \left[ P(\tilde C,t)-P(-\tilde C,t)\right] \ln\frac{P(\tilde C,t)}{P(-\tilde C,t)} \geq 0
\label{entrprod}
\ee
in terms of the probability density $P(\tilde C,t)$ and this expression is always non-negative because $(p-q)\ln(p/q)\geq 0$ for any non-vanishing real numbers $p$ and $q$.  Using the fluctuation theorem~(\ref{FT-tot-current}), the expression~(\ref{entrprod}) gives the dissipated power divided by the thermal energy
\be
\frac{1}{k_{\rm B}}\frac{d_{\rm i}S}{dt}= \lim_{t\to\infty}\frac{1}{t} \, A \langle\tilde C\rangle_t= \frac{VI(V)}{k_{\rm B}T} \geq 0 \, ,
\ee
as expected. Hence, the entropy production is non-negative in accord with the second law of thermodynamics.

\section{Conclusion and perspectives}
\label{Sec-concl}

In this paper, a spatially extended stochastic description has been introduced for the transport of electrons and holes in semiconducting diodes.  The description is based on stochastic partial differential equations for the charge carrier densities.  These balance equations are coupled to the Poisson equation for the electric potential generated by the charges.  The description is consistent with the laws of electricity and thermodynamics.  In the noiseless limit, the macroscopic description of the diode is recovered \cite{AM76,S49}. The spatially extended description allows us to define the measured electric current by including the contribution of the displacement current, which is essential for the study of current fluctuations \cite{BB00,AG09}.

For the purpose of simulating numerically the stochastic process, space is discretized into small cells.  In Appendix~\ref{AppA}, a Markov jump process is introduced, which is ruled by a master equation for the time evolution of the probability that the cells contain given numbers of electrons and holes.  If these numbers are large enough, the Markov jump process can be replaced by a Langevin stochastic process involving Gaussian white noises, as shown in Appendix~\ref{AppB}.  The numerical simulation of the Langevin stochastic process is significantly more efficient than the one of the Markov jump process, although giving results of comparable accuracy for large numbers of carriers in the cells.  The profiles of carrier densities and electric potential obtained with this stochastic algorithm agree with those calculated with the macroscopic mean-field equations under equilibrium and nonequilibrium stationary conditions.  In this way, current-voltage characteristics are computed for different concentration ratios of majority to minority carriers.  The Shockley model for the $I$-$V$ characteristic of the diode is shown to be valid under the extreme condition where the concentration is overwhelmingly larger for the majority carriers than for the minority carriers.

Since the stochastic description satisfies local detailed balance in consistency with microreversibility, the fluctuation theorem holds for the carrier current and the measured total current, as shown in Sec.~\ref{Sec-FT}. The macroscopic affinity given by the applied voltage is reached asymptotically in time as predicted by the fluctuation theorem.  The convergence to the macroscopic affinity is remarkably faster for the total current than the carrier current. The reason is that the inclusion of the displacement current in the total current expresses the effects of the long-ranged Coulomb interaction on the measurement of the current fluctuations. Therefore, the random jumps of the charge carriers anywhere inside the diode have an instantaneous effect on the measured total current in the quasi-static limit of Maxwell's equations. In this regard, the rapid convergence to the macroscopic affinity given by the applied voltage justifies the description of electronic devices in terms of global current-voltage characteristics.

To conclude, our study shows that the fluctuation theorem plays a fundamental role in diodes and that the experimental investigation of the fluctuation theorem in diodes could test its validity in highly nonlinear response regimes.  Moreover, the spatially extended stochastic processes we have here developed provide a powerful computational tool for the simulation of electronic devices in semiconductor technology.

\vskip 1 cm

\section*{Acknowledgments}

Financial support from the China Scholarship Council, the Universit\'e libre de Bruxelles (ULB), and the Fonds de la Recherche Scientifique~-~FNRS under the Grant PDR~T.0094.16 for the project ``SYMSTATPHYS" is acknowledged.


\appendix
\section{Discretized Markov jump process}
\label{AppA}

\subsection{Master equation of the process}

At the mesoscopic level, the evolution of electron and hole distributions in the diode can be successfully described as a Markov jump process, which is formulated in terms of a master equation. The fluctuations down to the mesoscopic scale can be fully characterized by such a process.

The stochastic model we here introduce incorporates the self-consistent electric field, which is generated by the fluctuating distribution of charges. The diode is spatially discretized by dividing it into cells, each with the same volume~$\Omega$ and same length $\Delta x$. The cross section is thus given by $\Sigma=\Omega/\Delta x$.  Accordingly, there is a total of $L=l/\Delta x$ cells indexed $i=1,2,\dots,L$. The indexes $i=0$ and~$i=L+1$ are respectively used to refer to the left and right reservoirs, which impose certain boundary conditions to the diode. The numbers of electrons, holes, acceptors, and donors in each cell are respectively given by $N_{i}=n(x_{i})\Omega$, $P_{i}=p(x_{i})\Omega$, $A_{i}=a(x_{i})\Omega$, and $D_{i}=d(x_{i})\Omega$, with $x_{i}=(i-0.5)\Delta x$ ($i=1,2,\dots,L$). The numbers of electrons and holes in the reservoirs ($N_{i}$ and $P_{i}$ for $i=0$ and $i=L+1$) are maintained constant in time. 

\begin{widetext}

The system state is specified by the electron numbers ${\bf N}=(N_{i})_{i=1}^L$ and the hole numbers ${\bf P}=(P_{i})_{i=1}^L$ in the cells and they evolve in time according to the network:
\begin{equation}
\begin{array}{ccccccccccccc}
\bar{N}_{\rm L} & \autorightleftharpoons{$\scriptstyle W_0^{(+N)}$}{$\scriptstyle W_0^{(-N)}$} & N_1 & \autorightleftharpoons{$\scriptstyle W_1^{(+N)}$}{$\scriptstyle W_1^{(-N)}$}& N_2 & \autorightleftharpoons{$\scriptstyle W_2^{(+N)}$}{$\scriptstyle W_2^{(-N)}$} & \cdots & \autorightleftharpoons{$\scriptstyle W_{L-2}^{(+N)}$}{$\scriptstyle W_{L-2}^{(-N)}$} & N_{L-1}& \autorightleftharpoons{$\scriptstyle W_{L-1}^{(+N)}$}{$\scriptstyle W_{L-1}^{(-N)}$} & N_L & \autorightleftharpoons{$\scriptstyle W_L^{(+N)}$}{$\scriptstyle W_L^{(-N)}$} & \bar{N}_{\rm R} \\
 & & {\scriptstyle W_1^{(+)}}\updownarrow{\scriptstyle W_1^{(-)}} & & {\scriptstyle W_2^{(+)}}\updownarrow{\scriptstyle W_2^{(-)}} & & \cdots & &{\scriptstyle W_{L-1}^{(+)}}\updownarrow{\scriptstyle W_{L-1}^{(-)}} & & {\scriptstyle W_L^{(+)}}\updownarrow{\scriptstyle W_L^{(-)}} & &  \\
\bar{P}_{\rm L} & \autorightleftharpoons{$\scriptstyle W_0^{(+P)}$}{$\scriptstyle W_0^{(-P)}$} &P_1 & \autorightleftharpoons{$\scriptstyle W_1^{(+P)}$}{$\scriptstyle W_1^{(-P)}$} & P_2 & \autorightleftharpoons{$\scriptstyle W_2^{(+P)}$}{$\scriptstyle W_2^{(-P)}$} & \cdots & \autorightleftharpoons{$\scriptstyle W_{L-2}^{(+P)}$}{$\scriptstyle W_{L-2}^{(-P)}$} & P_{L-1} & \autorightleftharpoons{$\scriptstyle W_{L-1}^{(+P)}$}{$\scriptstyle W_{L-1}^{(-P)}$} & P_L & \autorightleftharpoons{$\scriptstyle W_L^{(+P)}$}{$\scriptstyle W_L^{(-P)}$} & \bar{P}_{\rm R} \nonumber
\end{array}
\end{equation}
The probability ${\cal P}({\bf N},{\bf P};t)$ that the cells contain the particle numbers ${\bf N}$ and ${\bf P}$ for time $t$ is ruled by the master equation
\bea\label{eq_master_equation}
\frac{d{\cal P}}{dt}&=&\sum_{i=0}^{L} \Biggl[\left({\rm e}^{+\partial_{N_i}}{\rm e}^{-\partial_{N_{i+1}}}-1\right)W_i^{(+N)}{\cal P}+\left({\rm e}^{-\partial_{N_i}}{\rm e}^{+\partial_{N_{i+1}}}-1\right)W_i^{(-N)}{\cal P} \nonumber\\
&&\qquad+\left({\rm e}^{+\partial_{P_i}}{\rm e}^{-\partial_{P_{i+1}}}-1\right)W_i^{(+P)}{\cal P}+\left({\rm e}^{-\partial_{P_i}}{\rm e}^{+\partial_{P_{i+1}}}-1\right)W_i^{(-P)}{\cal P}\Biggr] \nonumber\\
&&+\sum_{i=1}^{L}\Biggl[\left({\rm e}^{-\partial_{N_i}}{\rm e}^{-\partial_{P_i}}-1\right)W_i^{(+)}{\cal P}+\left({\rm e}^{+\partial_{N_i}}{\rm e}^{+\partial_{P_i}}-1\right)W_i^{(-)}{\cal P}\Biggr] \text{.}
\eea
with the transition rates
\bea
&& W_i^{(+N)}=\frac{D_n}{\Delta x^2}\psi(\Delta U_{i, i+1}^{(N)})N_i , \label{W+N}\\ 
&& W_i^{(-N)}=\frac{D_n}{\Delta x^2}\psi(\Delta U_{i+1, i}^{(N)})N_{i+1} , \label{W-N}\\ 
&& W_i^{(+P)}=\frac{D_p}{\Delta x^2}\psi(\Delta U_{i, i+1}^{(P)})P_i , \label{W+P}\\
&& W_i^{(-P)}=\frac{D_p}{\Delta x^2}\psi(\Delta U_{i+1, i}^{(P)})P_{i+1} , \label{W-P}\\
&& W_i^{(+)}=\Omega k_+ , \label{W+R}\\
&& W_i^{(-)}=\Omega k_- \frac{N_i}{\Omega}\frac{P_i}{\Omega} , \label{W-R}
\eea
where $U$ denotes the electrostatic energy stored in the system and $\Delta U_{i, i+1}$ the energy difference associated with the exchange of one particle between the cells $i$ and $i+1$. The function $\psi(\Delta U)$ is defined as
\be
\psi(\Delta U)=\frac{\beta\Delta U}{\exp(\beta\Delta U)-1} ,
\ee
which satisfies the condition
\be
\psi(\Delta U)=\psi(-\Delta U)\exp(-\beta\Delta U) ,
\label{detbal}
\ee
guaranteeing detailed balance at equilibrium.

At the ends of the chain, we have that $\exp(\pm\partial_X)=1$ for $X=N_0$, $P_0$, $N_{L+1}$, and $P_{L+1}$ in the master equation~(\ref{eq_master_equation}).  Indeed, the quantities $N_0=\bar{N}_{\rm L}$, $P_0=\bar{P}_{\rm L}$, $N_{L+1}=\bar{N}_{\rm R}$, and $P_{L+1}=\bar{P}_{\rm R}$ are associated with the reservoirs at the boundaries and they thus take constant values.  These considerations determine the transition rates at the boundaries.

\subsection{Discretized Poisson equation}

The Poisson equation is replaced by its discretized version
\be
\frac{\phi_{i+1}-2\phi_{i}+\phi_{i-1}}{\Delta x^{2}}=-\frac{e}{\epsilon\Omega}(P_{i}-N_{i}+D_{i}-A_{i})
\ee
with the boundary conditions $\phi_{0}=\phi_{\rm L}$ and $\phi_{L+1}=\phi_{\rm R}$. This linear system should be solved every time a particle transition occurs. It can be written in the matrix form
\be
\boldsymbol{\mathsf C}\cdot \pmb{\phi}={\bf Z} 
\ee
with the set of electric potentials $\pmb{\phi}=(\phi_{i})_{i=1}^L$ in the cells, the $L\times L$ symmetric matrix $\boldsymbol{\mathsf C}$ composed of the elements
\be
({\boldsymbol{\mathsf C}})_{ij}=\gamma \left( -\delta_{i+1,j}+2\delta_{i,j}-\delta_{i-1,j}\right) \qquad\mbox{with}\qquad \gamma=\frac{\epsilon\Omega}{\Delta x^{2}}
\ee
for $i,j=1,2,\dots,L$, and
\be
{\bf Z}=e\left(P_i-N_i+D_i-A_i\right)_{i=1}^L+\gamma(\phi_{\rm L}, 0, \dots, 0, \phi_{\rm R}) .
\label{Z}
\ee
The electric potential is thus given by
\begin{align}
\pmb{\phi}={\pmb{\mathsf C}}^{-1}\cdot{\bf Z}
\end{align}
with
\begin{align}
\Big({\boldsymbol{\mathsf C}}^{-1}\Big)_{ij}=
\begin{cases}
\frac{i}{\gamma(L+1)}(L+1-j) & \text{if}  \quad i\leq j \text{,} \\
\frac{j}{\gamma(L+1)}(L+1-i) & \text{if}  \quad i>j \text{.}
\end{cases}
\label{eq_inverse_C}
\end{align}

\subsection{Discretized electrostatic energy}

The electrostatic energy  can be expressed as
\be
U(\pmb{\phi})=\frac{1}{2}\, \pmb{\phi} \cdot \boldsymbol{\mathsf C}\cdot \pmb{\phi}
\ee
in terms of the electric potential, or equivalently as
\be
U({\bf Z})=\frac{1}{2}\, {\bf Z} \cdot \boldsymbol{\mathsf C}^{-1}\cdot {\bf Z}
\ee
in terms of the charges~(\ref{Z}).  Therefore, the change of electrostatic energy during the transition of an electron of charge $-e$ from the $i^{\rm th}$ to the $(i+1)^{\rm th}$ cell is given by
\be
\Delta U_{i,i+1}^{(N)}=U({\bf Z'})-U({\bf Z}) = \frac{1}{2}\left({\bf Z'} \cdot {\boldsymbol{\mathsf C}} ^{-1} \cdot {\bf Z'} - {\bf Z} \cdot {\boldsymbol{\mathsf C}}^{-1} \cdot{\bf Z}\right)
\ee
with
\be
Z'_j = Z_j+e\delta_{i,j}-e\delta_{i+1,j} .
\ee
Introducing the notations ${\bf n}_i$ and ${\bf p}_i$ such that 
\be
({\bf n}_i)_j=-e\delta_{i,j} \qquad \mbox{and} \qquad ({\bf p}_i)_j=+e\delta_{i,j} ,
\label{ni-pi}
\ee
we thus have that
\bea
&&\Delta U_{i,i+1}^{(N)}=U({\bf Z}+{\bf n}_{i+1}-{\bf n}_i)-U({\bf Z})=-e(\phi_{i+1}-\phi_i)+\frac{e^2}{2}\Big[\left(\boldsymbol{\mathsf C}^{-1}\right)_{i,i}-2\left({\boldsymbol{\mathsf C}}^{-1}\right)_{i,i+1}+\left({\boldsymbol{\mathsf C}}^{-1}\right)_{i+1,i+1}\Big] , \\
&&\Delta U_{i,i+1}^{(P)}=U({\bf Z}+{\bf p}_{i+1}-{\bf p}_i)-U({\bf Z})=+e(\phi_{i+1}-\phi_i)+\frac{e^2}{2}\Big[\left(\boldsymbol{\mathsf C}^{-1}\right)_{i,i}-2\left({\boldsymbol{\mathsf C}}^{-1}\right)_{i,i+1}+\left({\boldsymbol{\mathsf C}}^{-1}\right)_{i+1,i+1}\Big] \text{.}
\eea
Using Eq.~(\ref{eq_inverse_C}), we find that
\bea
&&\Delta U_{i,i+1}^{(N)}=-e(\phi_{i+1}-\phi_i)+\frac{e^2L\Delta x^2}{2(L+1)\epsilon\Omega} \text{,} \label{dUN}\\
&&\Delta U_{i,i+1}^{(P)}=+e(\phi_{i+1}-\phi_i)+\frac{e^2L\Delta x^2}{2(L+1)\epsilon\Omega} \text{.} \label{dUP}
\eea
We notice that, for transitions at the boundaries, these expressions hold by taking the values of the potentials in the reservoirs, $\phi_0=\phi_{\rm L}$ and $\phi_{L+1}=\phi_{\rm R}$.

\subsection{Cyclic paths and their affinity}
\label{AppA4}

According to the network theory of Markov jump processes \cite{S76}, a graph $G$ can be associated with the master equation in such a way that each state of the system corresponds to a vertex and the different allowed transitions $\omega\rightleftharpoons\omega'$ between the states are represented by edges. In the so-constructed graph, cyclic paths are sequences of edges joining a finite set of vertices and coming back to the starting vertex.  Denoting by $\omega$ the vertices and $e$ the edges of the graph, the affinity of the cyclic path $\cal C$ is defined as
\be
A({\cal C}) \equiv \ln \prod_{e\in{\cal C}} \frac{W(\omega\stackrel{e}{\rightarrow}\omega')}{W(\omega\stackrel{e}{\leftarrow}\omega')}
\label{Aff-dfn}
\ee
in terms of the ratio of transition rates along the path divided by the transition rates along the reversed path \cite{S76}. This affinity characterizes the nonequilibrium constraints imposed by the boundaries on the cyclic path.  Let us consider several examples of cyclic paths.

A first example of cyclic path is the transfer of a hole from the left to the right reservoir:
\be
{\cal C}_1: \qquad  {\bf Z} \stackrel{W_0^{(+P)}}{\rightarrow} {\bf Z}+{\bf p}_1 \stackrel{W_1^{(+P)}}{\rightarrow} {\bf Z}+{\bf p}_2 \rightarrow \cdots \rightarrow {\bf Z}+{\bf p}_{L-1}\stackrel{W_{L-1}^{(+P)}}{\rightarrow} {\bf Z}+{\bf p}_L \stackrel{W_{L}^{(+P)}}{\rightarrow} {\bf Z} ,
\ee  
written in terms of the charges~(\ref{Z}) and the notation~(\ref{ni-pi}).
The corresponding nonequilibrium constraint is determined by the applied voltage that we should recover by calculating the affinity.  Indeed, the transition rates $W_i^{(+P)}$ defined by Eq.~(\ref{W+P}) involve the energy differences
\be
\Delta U_{i,i+1}^{(P)} = U({\bf Z}+{\bf p}_{i+1})-U({\bf Z}+{\bf p}_i) = e(\phi_{i+1}-\phi_i) + \frac{e^2(L-2i)}{2\gamma(L+1)} =-\Delta U_{i+1,i}^{(P)} 
\label{DU-1}
\ee
for $i=0,1,\dots,L$ with $\phi_0=\phi_{\rm L}$ and $\phi_{L+1}=\phi_{\rm R}$.  Substituting Eqs.~(\ref{W+P}) and~(\ref{W-P}) in the definition~(\ref{Aff-dfn}) and using the detailed balance relation~(\ref{detbal}), we find that
\be
A({\cal C}_1) = \ln\left[ \frac{p_{\rm L}}{p_{\rm R}} \, {\rm e}^{\beta e (\phi_{\rm L}-\phi_{\rm R})}\right] = \beta e V
\ee
according to Eq.~(\ref{voltage}), which shows the consistency of the scheme for this cyclic path.

The same affinity should be obtained for the second example where an electron is transferred from the right to the left reservoir
\be
{\cal C}_2: \qquad  {\bf Z} \stackrel{W_L^{(-N)}}{\rightarrow} {\bf Z}+{\bf n}_L \stackrel{W_{L-1}^{(-N)}}{\rightarrow} {\bf Z}+{\bf n}_{L-1} \rightarrow \cdots \rightarrow {\bf Z}+{\bf n}_2\stackrel{W_1^{(-N)}}{\rightarrow} {\bf Z}+{\bf n}_1 \stackrel{W_{0}^{(-N)}}{\rightarrow} {\bf Z} .
\ee  
Here, the transition rates $W_i^{(-N)}$ defined by Eq.~(\ref{W-N}) are determined by the energy differences
\be
\Delta U_{i+1,i}^{(N)} = U({\bf Z}+{\bf n}_i)-U({\bf Z}+{\bf n}_{i+1}) = e(\phi_{i+1}-\phi_i) - \frac{e^2(L-2i)}{2\gamma(L+1)} =-\Delta U_{i,i+1}^{(N)} ,
\ee
giving the affinity
\be
A({\cal C}_2) = \ln\left[ \frac{n_{\rm R}}{n_{\rm L}} \, {\rm e}^{\beta e (\phi_{\rm L}-\phi_{\rm R})}\right] = \beta e V
\ee
with Eq.~(\ref{voltage}), which confirms the expectation.

As a third example, we consider the cyclic path where a hole moves from the left reservoir to the $j^{\rm th}$ cell, followed by the transfer of an electron from the right reservoir also to the $j^{\rm th}$ cell, where they recombine:
\be
{\cal C}_3: \qquad  {\bf Z} \stackrel{W_0^{(+P)}}{\rightarrow} {\bf Z}+{\bf p}_1 \stackrel{W_1^{(+P)}}{\rightarrow}\cdots \stackrel{W_{j-1}^{(+P)}}{\rightarrow} {\bf Z}+{\bf p}_j\stackrel{W_{L}^{(-N)}}{\rightarrow} {\bf Z}+{\bf p}_j+{\bf n}_L \stackrel{W_{L-1}^{(-N)}}{\rightarrow}\cdots \stackrel{W_{j}^{(-N)}}{\rightarrow}{\bf Z}+{\bf p}_j+{\bf n}_j \stackrel{W_{j}^{(-)}}{\rightarrow}{\bf Z} .
\ee
Here, the transition rates involve the energy differences~(\ref{DU-1}) for the transitions of the hole and
\be
\Delta U_{i+1,i}^{(N)} = U({\bf Z}+{\bf p}_j+{\bf n}_i)-U({\bf Z}+{\bf p}_j+{\bf n}_{i+1}) = e(\phi_{i+1}-\phi_i) - \frac{e^2(L-2i+2j)}{2\gamma(L+1)} =-\Delta U_{i,i+1}^{(N)}
\ee
for the transitions of the electron in the presence of the hole in the $j^{\rm th}$ cell.  Substituting the corresponding transition rates given by Eqs.~(\ref{W+N})-(\ref{W-R}) in the definition~(\ref{Aff-dfn}) and using the detailed balance relation~(\ref{detbal}), we here get
\be
A({\cal C}_3) = \ln\left[ \frac{k_-}{k_+} \, p_{\rm L}\, n_{\rm R} \, {\rm e}^{\beta e (\phi_{\rm L}-\phi_{\rm R})}\right] = \beta e V .
\ee
Since $k_+=k_- n_{\rm L}p_{\rm L}=k_- n_{\rm R}p_{\rm R}$, we again find the applied voltage~(\ref{voltage}), as it should.

Similar results can be obtained for other cyclic paths.

\section{Langevin stochastic process}
\label{AppB}

\subsection{Discretized stochastic equations}

In the limit of large particle numbers in the cells (i.e., $N_i\gg 1$ and $P_i\gg 1$ for all indexes $i$), the Markov jump process can be approximated by a Langevin stochastic process.  In this approximation, the operators $\exp(\pm\partial_X)$ are replaced by their expansion $1\pm\partial_X+\frac{1}{2}\partial_X^2\pm\cdots$ in Eq.~(\ref{eq_master_equation}) keeping only the terms up to second partial derivatives, so that we get the following equation
\bea
\partial_t{\mathscr P}&=&\sum_{i=0}^{L}\Bigg\{-\partial_{N_i}\left[\left(W_{i-1}^{(+N)}-W_{i-1}^{(-N)}-W_{i}^{(+N)}+W_{i}^{(-N)}\right){\mathscr P}\right] \nonumber\\
&&+\partial_{N_i}^2\left[\frac{1}{2}\left(W_{i-1}^{(+N)}+W_{i-1}^{(-N)}+W_{i}^{(+N)}+W_{i}^{(-N)}\right){\mathscr P}\right] \nonumber\\
&&+\partial_{N_i}\partial_{N_{i+1}}\left[-\left(W_{i}^{(+N)}+W_{i}^{(-N)}\right){\mathscr P} \right]+(N\rightleftharpoons P)\Bigg\}\nonumber\\
&&+\sum_{i=1}^{L}\Bigg\{-\left(\partial_{N_i}+\partial_{P_i}\right)\left[\left(W_{i}^{(+)}-W_{i}^{(-)}\right){\mathscr P}\right]+\left(\partial_{N_i}+\partial_{P_i}\right)^2\left[\frac{1}{2}\left(W_{i}^{(+)}+W_{i}^{(-)}\right){\mathscr P}\right]\Bigg\}
\eea
for the time evolution of the probability density ${\mathscr P}({\bf N},{\bf P};t)$ \cite{G05}.
This shows that the variables $N_i$ and $P_i$ obeys the following stochastic differential equations of Langevin type
\bea
&& \frac{dN_i}{dt}=F_{i-1}^{(N)}-F_{i}^{(N)}+R_i , \label{dNdt}\\
&& \frac{dP_i}{dt}=F_{i-1}^{(P)}-F_{i}^{(P)}+R_i , \label{dPdt}
\eea
expressed in terms of the fluxes and reaction rates
\bea
&& F_i^{(N)}=W_i^{(+N)}-W_i^{(-N)}+\sqrt{W_i^{(+N)}+W_i^{(-N)}}\xi_i^{(N)}(t) , \label{FiN}\\
&& F_i^{(P)}=W_i^{(+P)}-W_i^{(-P)}+\sqrt{W_i^{(+P)}+W_i^{(-P)}}\xi_i^{(P)}(t) , \label{FiP}\\
&& R_i=W_i^{(+)}-W_i^{(-)}+\sqrt{W_i^{(+)}+W_i^{(-)}}\eta_i(t) , \label{Ri}
\eea
and the Gaussian white noises
\bea
&& \langle\xi_i^{(N)}(t)\rangle=\langle\xi_i^{(P)}(t)\rangle=\langle\eta_i(t)\rangle=0 , \label{xieta1}\\
&& \langle\xi_i^{(N)}(t)\xi_j^{(N)}(t')\rangle=\langle\xi_i^{(P)}(t)\xi_j^{(P)}(t')\rangle=\langle\eta_i(t)\eta_j(t')\rangle=\delta_{ij}\delta(t-t') , \label{xieta2}\\
&& \langle\xi_i^{(N)}(t)\xi_j^{(P)}(t')\rangle=\langle\eta(t)\xi_j^{(N)}(t')\rangle= \langle\eta(t)\xi_j^{(P)}(t')\rangle=0 . \label{xieta2c}
\eea
The boundary conditions are imposed at the left and right reservoirs by setting $N_0=\bar{N}_{\rm L}$, $P_0=\bar{P}_{\rm L}$, $\phi_0=\phi_{\rm L}$, $N_{L+1}=\bar{N}_{\rm R}$, $P_{L+1}=\bar{P}_{\rm R}$, and $\phi_{L+1}=\phi_{\rm R}$ in the transition rates.

To implement numerically the Langevin stochastic equations~(\ref{dNdt})-(\ref{dPdt}), time is discretized into equal intervals $\Delta t$ and the white noises are replaced by independent identically distributed normal random variables.

\subsection{The continuum limit}

We can recover the stochastic partial differential equations~(\ref{dndt-c})-(\ref{dpdt-c}) with the current and rate densities~(\ref{jn})-(\ref{sigma})  from the Langevin stochastic equations~(\ref{dNdt})-(\ref{Ri}), as follows.  First, we note that the approximation $\psi(\Delta U) \simeq \exp( -\beta\Delta U/2)$ holds if $\beta\Delta U\ll 1$.  Next, using Eqs.~(\ref{W+N}), (\ref{W-N}), and~(\ref{dUN}) in the limit $\Delta x\to 0$, Eq.~(\ref{FiN}) gives the flux
\be
F_i^{(N)} \simeq - \frac{D_n}{\Delta x^2} \, {\rm e}^{\beta e\phi_{i+1/2}}\left({\rm e}^{-\beta e\phi_{i+1}} \, N_{i+1}-{\rm e}^{-\beta e\phi_{i}} \, N_{i}\right) 
+\frac{1}{\Delta x} \sqrt{D_n(N_i+N_{i+1})} \ \xi_i^{(N)}(t) \, , 
\ee
where $\phi_{i+1/2} \simeq (\phi_i + \phi_{i+1})/2$.  Besides, using Eqs.~(\ref{W+R}) and~(\ref{W-R}), the rate~(\ref{Ri}) becomes
\be
R_i=\Omega\left(k_+-k_-\frac{N_i}{\Omega}\frac{P_i}{\Omega}\right) + \sqrt{\Omega\left(k_+ + k_-\frac{N_i}{\Omega}\frac{P_i}{\Omega}\right)} \ \eta_i(t) \, .
\ee
Substituting these expressions into Eq.~(\ref{dNdt}) and dividing it by $\Omega$, we find the electron balance equation~(\ref{dndt-c}) with the current density~(\ref{jn}) in the form~(\ref{jn-exp}), together with the source~(\ref{sigma}).  The hole balance equation~(\ref{dpdt-c}) is similarly deduced from Eq.~(\ref{dPdt}).  Because of Eqs.~(\ref{FiN})-(\ref{xieta2c}), the noise fields obey Eqs.~(\ref{aj})-(\ref{Gss}) since $\delta_{ij}/\Omega\to \delta^3({\bf r}-{\bf r}')$ in the limit $\Omega\to 0$.  The stochastic partial differential equations are thus recovered in the continuum limit.

\subsection{The currents}

In the framework of the Langevin stochastic process, the net charge current~(\ref{current}) at the location $x=i\Delta x$ is approximately given by
\be
{\cal I}(t) = e \int d\pmb{\Sigma}\cdot({\bf j}_p-{\bf j}_n)\simeq e\left( F_i^{(P)}-F_i^{(N)}\right),
\label{current-discrete}
\ee
in terms of the fluxes~(\ref{FiN})-(\ref{FiP}).  The current that is experimentally measured is the total current~(\ref{tot-current}), which includes the contribution from the displacement current.
After spatial discretization and surface integration over the section area $\Sigma=\Omega/\Delta x$, this contribution becomes
\be
\Sigma\epsilon\partial_t{\cal E}_{x,i}\simeq -\frac{e\epsilon\Sigma}{\Delta x} \sum_{j=0}^{L}\left[ ({\boldsymbol{\mathsf C}}^{-1})_{i+1,j+1}-({\boldsymbol{\mathsf C}}^{-1})_{i+1,j}
-({\boldsymbol{\mathsf C}}^{-1})_{i,j+1}
+({\boldsymbol{\mathsf C}}^{-1})_{i,j}\right](F_j^{(P)}-F_j^{(N)}),
\ee
where
\be
({\boldsymbol{\mathsf C}}^{-1})_{i+1,j+1}-({\boldsymbol{\mathsf C}}^{-1})_{i+1,j}
-({\boldsymbol{\mathsf C}}^{-1})_{i,j+1}
+({\boldsymbol{\mathsf C}}^{-1})_{i,j}=\frac{L}{\gamma(L+1)}\, \delta_{ij} - \frac{1}{\gamma(L+1)}\, (1-\delta_{ij}) .
\ee
Therefore, the discretized form of the total current is given by
\be
\tilde{\cal I}(t)= \int d\pmb{\Sigma}\cdot\left[ e({\bf j}_p-{\bf j}_n)+\epsilon\, \partial_t\pmb{\cal E}\right]  \simeq \frac{e}{L+1} \sum_{j=0}^{L} \left( F_j^{(P)}-F_j^{(N)}\right) ,
\label{tot-current-Langevin}
\ee
which is independent of the location $x$. We notice that the expression~(\ref{tot-current-Langevin}) can also be obtained using the Ramo-Shockley theorem \cite{S38,R39,AG09}.

\end{widetext}



\begin{thebibliography}{99}

\bibitem{AM76} N. W. Ashcroft and N. D. Mermin, {\it Solid State Physics} (Saunders College, Philadelphia, 1976).

\bibitem{S49} W. Shockley, Bell Syst. Tech. J. {\bf 28}, 33 (1949).

\bibitem{O31a} L. Onsager, Phys. Rev. {\bf 37}, 405 (1931). 

\bibitem{O31b} L. Onsager, Phys. Rev. {\bf 38}, 2265 (1931). 

\bibitem{C45} H. B. G. Casimir, Rev. Mod. Phys. {\bf 17}, 343 (1945).

\bibitem{G96} G. Gallavotti, Phys. Rev. Lett. {\bf 77}, 4334 (1996).

\bibitem{ES02} D. J. Evans and D. J. Searles, Adv. Phys. {\bf 51}, 1529 (2002).

\bibitem{EHM09} M.~Esposito, U.~Harbola, and S.~Mukamel, Rev. Mod. Phys. {\bf 81}, 1665 (2009).

\bibitem{CHT11} M. Campisi, P. H\"anggi, and P. Talkner, Rev. Mod. Phys. {\bf 83}, 771 (2011); {\it Erratum}, {\it ibid.} {\bf 83}, 1653 (2011).

\bibitem{J11} C. Jarzynski, Annu. Rev. Condens. Matter Phys. {\bf 2}, 329 (2011).

\bibitem{S12} U. Seifert, Rep. Prog. Phys. {\bf 75}, 126001 (2012).

\bibitem{G13} P. Gaspard, New J. Phys. {\bf 15}, 115014 (2013).

\bibitem{J28} J. B. Johnson, Phys. Rev. {\bf 32}, 97 (1928).

\bibitem{N28} H. Nyquist, Phys. Rev. {\bf 32}, 110 (1928).

\bibitem{BB00} Ya. M. Blanter and M. B\"uttiker, Phys. Rep. {\bf 336}, 1 (2000).

\bibitem{AG04} D. Andrieux and P. Gaspard, J. Chem. Phys. {\bf 121}, 6167 (2004).

\bibitem{AG07JSM} D.~Andrieux and P.~Gaspard, J. Stat. Mech.: Th. Exp. P02006 (2007).

\bibitem{HPPG11} P. I. Hurtado, C. P\'erez-Espigares, J. J. del Pozo, and P. L. Garrido, Proc. Natl. Acad. Sci. (USA) {\bf 108}, 7704 (2011).

\bibitem{vZCC04} R. van Zon, S. Ciliberto, and E. G. D. Cohen, Phys. Rev. Lett. {\bf 92}, 130601 (2004).

\bibitem{GC05} N. Garnier and S. Ciliberto, Phys. Rev. E {\bf 71}, 060101 (2005).

\bibitem{JGC08} S. Joubaud, N. B. Garnier, and S. Ciliberto, Europhys. Lett. {\bf 82} 30007 (2008).

\bibitem{AG09} D. Andrieux and P. Gaspard, J. Stat. Mech. P02057 (2009).

\bibitem{H68} H. Hurwitz, Phys. Rev. {\bf 172}, 207 (1968).

\bibitem{LH10} J. Liu and J. He, Phys. Rev. E {\bf 82}, 022101 (2010).

\bibitem{AWBMJ91} M. Amman, R. Wilkins, E. Ben-Jacob, P. D. Maker, and R. C. Jaklevic, Phys. Rev. B {\bf 43}, 1146 (1991).

\bibitem{AG06} D. Andrieux and P. Gaspard, J. Stat. Mech. P01011 (2006).

\bibitem{J99} J. D. Jackson, {\it Classical Electrodynamics}, 3rd edition (Wiley, Hoboken, 1999).

\bibitem{G76} D. T. Gillespie, J. Comput. Phys. {\bf 22}, 403 (1976).

\bibitem{S76} J. Schnakenberg, Rev. Mod. Phys. {\bf 48}, 571 (1978).

\bibitem{AG07} D. Andrieux and P. Gaspard, J. Stat. Phys. {\bf 127}, 107 (2007).

\bibitem{B95} P. Billingsley, {\it Probability and Measure}, 3rd ed. (Wiley, New York, 1995).

\bibitem{G05} P. Gaspard, New J. Phys. {\bf 7}, 77 (2005).

\bibitem{S38} W. Shockley, J. Appl. Phys. {\bf 9}, 635 (1938).

\bibitem{R39} S. Ramo, Proc. IRE {\bf 27}, 584 (1939).

\end{thebibliography}
\end{document}